\font\rm=cmr12
\font\smallrm=cmr10
\font\it=cmti12
\font\bf=cmbx12
\font\smallit=cmti10
\baselineskip=18pt
\rm

\noindent \hfil {\bf {Normal Biconformal Spaces}}
\bigskip
\noindent \hfil James T. Wheeler

\noindent \hfil {\it {Department of Physics, Utah State University, Logan,
UT 84322}}

\noindent \hfil jwheeler@cc.usu.edu
\bigskip
\noindent \hfil Abstract
\smallskip

A new 8-dimensional conformal gauging avoids the unphysical size change,
third order gravitational field equations, and auxiliary fields that
prevent taking the conformal group as a fundamental symmetry.  We give the
structure equations, gauge transformations and intrinsic metric structure
for the new {\it {biconformal spaces}}.  We prove that a torsion-free
biconformal space with exact Weyl form, closed dilational curvature and
trace-free spacetime curvature admits a sub-bundle of vanishing Weyl form
homeomorphic to the Whitney sum bundle of the tangent bundle and the bundle
of orthonormal Lorentz frames over 4-dimensional spacetime.  Conversely,
any 4-dimensional spacetime extends uniquely to such a {\it {normal}}
biconformal space.
The Einstein equation holds iff the biconformal basis is orthonormal.
Unconstrained antisymmetric trace of the spacetime curvature provides a
closed 2-form, independent of the Weyl vector, consistently interpretable
as the electromagnetic field. The trace of the spacetime co-torsion
decouples from gravitational sources and serves as electromagnetic source.

\vfil
\break

\noindent \hfil {\bf {Normal Biconformal Spaces}}
\medskip
\noindent \hfil James T. Wheeler

\noindent \hfil {\it {Department of Physics, Utah State University, Logan,
UT 84322}}

\noindent \hfil jwheeler@cc.usu.edu
\bigskip
\noindent 1.  Introduction
\smallskip

Three long-standing difficulties hinder use of the 15-dimensional conformal
group
as an underlying symmetry of spacetime:  the prediction of unphysical size
change, the cubic, third order form of the gravitational field equations,
and the auxiliary field problem.  These difficulties are puzzling, since
the conformal group is mathematically natural (it is a simple group) and
physically necessary (since we certainly can use local units if we choose).
We overcome these problems by using a new 8-dimensional gauging of the
conformal group.  We show that the resulting class of non-Riemannian
differential geometries has a well-defined subclass equivalent to the class
of 4-dimensional spacetimes.

The first of the three problems is that of unphysical size change in
scale-invariant
theories.  Since the first introduction of scale invariance by Weyl in 1918
[1-5], the
physical changes of size induced by a non-integrable dilational gauge
potential have
prevented identifying that gauge field with the electromagnetic potential
despite its having
the proper gauge dependence.  Ultimately London [8] made the first steps
toward
our current U(1) gauge theory of electromagnetism.  But while
electromagnetism is well
understood, the presence of size change still provides an obstruction to
the direct use of
conformal gauge theory.

The second difficulty with conformal gauge theory is the complexity of the
gravitational
field equations.  Szekeres [49] showed that the conformal equavalent of the
vacuum Einstein equation includes an expression cubic in the spinorial
components of the Weyl curvature tensor and third order in derivatives of
the metric. While this more complicated field equation poses a technical
problem rather than an objection in principle, it would be more satisfying
if a simple limit of a conformal gauge theory led directly to the Einstein
equation.  Of course, the third order equations can be avoided by using a
Weyl geometry instead, but then we lose the extra symmetry provided by the
conformal group.  Moreover, there has been no satisfactory interpretation
of the dilations in 4-dimensional Weyl geometry.

Finally, while conformal gauge theories in 4-dimensions provide eleven
gauge fields on
spacetime, Crispim Rom\~{a}o, Ferber and Freund [40, 41], and independently
Kaku, Townsend and Van Nieuwenhuizen [42] showed that four of them are
auxiliary - their field equations could be solved algebraically, allowing
the fields to be
eliminated from the problem (see also [38, 39], [43,44]).  Wheeler [32]
extended the result to any conformal theory quadratic in the curvatures.

We resolve all three of these difficulties by gauging the conformal group
over an 8-dimensional base space instead of the usual 4-dimensional
spacetime base space.  To emphasize the resulting differences from the
usual gauging, including the effects of the reduced fiber symmetry and the
presence of a duality relation which holds between two sets of
4-dimensional variables, we call the new space {\it {biconformal space}}.
Our central result is a set of necessary and sufficient conditions for the
resulting 8-dimensional base space with 7-dimensional fibers to be
homeomorphic to a 4-dimensional Riemannian spacetime with 11-dimensional
fibers.  This result leads immediately to a simple invariant
characterization of the Einstein equation in biconformal space, as well as
a new geometric candidate for the electromagnetic field.

It is easy to see that the new gauging immediately resolves all three
existing problems with conformal gauge theory.  The auxiliary field problem
evaporates because the four fields which would have been auxiliary instead
necessarily form half of the solder form on the 8-dimensional base space.
Since the solder form is required to give a basis, it is manifestly
impossible to eliminate any of the eight translational gauge fields.  The
problem of dilations is also solved, because despite the presence of
nonvanishing dilational curvature on the 8-dimensional space the embedded
{\it {Riemannian}} spacetime is dilation-free.  Finally, we will show how
the existence of an invariant equivalent to the Einstein equation hinges on
the reduced fiber size.  Instead of the 11-dimensional symmetry of the
usual conformal gauging, the 8-dimensional gauging has only the 7
symmetries of the homogeneous Weyl group.  For the class of {\it {normal
biconformal spaces}} this 7-dimensional fiber symmetry leaves the usual
Einstein equation both scale and Lorentz invariant.

Beyond a statement of the necessary and sufficient conditions above, we
find that the antisymmetric part of the spacetime projection of the
co-solder form (i.e., the extra four components of the solder form)
provides a new geometric candidate for the electromagnetic field.  Unlike
previous unifications of gravity and electromagnetism based on scale
symmetry [1-30], this identification is independent of the presence or
absence of the Weyl vector.  We identify the trace of the spacetime part of
the co-torsion as a phenomenological electromagnetic source, and show that
it decouples from gravitational sources.  While the presence of two
electromagnetic-type gauge fields (the Weyl vector and part of the new
co-torsion field) suggests the possibility of a geometric model for the
electroweak interaction, we suggest as yet no specific candidates for the
$W^{\pm}$ particles.

One further note on our gauging is in order.  Instead of basing our
arguments on a
particular globally conformally invariant action and introducing guage
fields to make the
conformal symmetry local, we will use standard fiber bundle techniques
[34-37, 45, 46] to build a mathematical structure consistent with {\it
{any}} action which can be built from the conformal curvatures and other
natural structures on the space.  This has the double advantage of greater
generality and of making the underlying mathematical structure clear.

The organization of the paper is as follows.  In Sec.(2) we define some
basic notation and
terminology.  Then in Sec.(3) we present the new gauging of the conformal
group,
including its structure equations, gauge transformations and metric
structure.   Sec.(4) consists of a lemma used in proving the central
theorem.  The lemma gives the consequences of vanishing torsion (but not
co-torsion) for the form of the connection.  Our central theorem, proved in
Sec.(5), states that the vanishing Weyl vector sub-bundle of a torsion-free
biconformal space with exact Weyl vector, closed dilation and trace-free
spacetime curvature is homeomorphic to the Whitney sum bundle of the
tangent and orthonormal frame bundles of a 4-dimensional Riemannian base
manifold.  Finally, in Sec. (6), we determine field equations for Einstein
gravity and Maxwell electromagnetism as geometric conditions in biconformal
spaces.

\medskip
\noindent 2.  Preliminaries
\medskip
Our notation for a Riemannian geometry is as follows.  Let ${\bf {e}}^{a} =
e_{\mu}^{\enskip a} {\bf {d}}x^{\mu}$ and  $\omega_{b}^{a}$ be the vierbein
and spin
connection 1-forms, respectively, of a 4-dimensional Riemannian spacetime
$({\cal {M}}, g)$, with metric $g_{\mu \nu} = \eta_{ab} e_{\mu}^{\enskip a}
e_{\nu}^{\enskip b}$ where $\eta_{ab} = diag(1,1,1,-1)$.  Let $e^{\enskip
\mu}_{a}$ denote the inverse of the matrix of components $e_{\mu}^{\enskip
a}$ of the vierbein.  Differential forms are denoted by boldface or greek
symbols, and the wedge product is always assumed between adjacent forms.
We use Greek indices for coordinate bases and Latin labels for orthonormal
bases.  The two sets of 1-forms $({\bf {e}}^{a}, \omega_{b}^{a})$ satisfy
the Maurer-Cartan structure equations
$$\eqalignno{
{\bf {de}}^{a} &= {\bf {e}}^{b} \omega_{b}^{a} &(2.1a)  \cr
{\bf {d}}\omega_{b}^{a} &= \omega_{b}^{c} \omega_{c}^{a} + {\bf {R}}_{b}^{a}
&(2.1b) \cr}$$
where ${\bf {R}}_{b}^{a} = {\scriptstyle {1 \over 2}} R_{\enskip bcd}^{a} {\bf
{e}}^{c}{\bf {e}}^{d}$ is the curvature 2-form.  Eqs.(2.1) describe a
connection on the
10-dimensional principal fiber bundle of orthonormal frames.  The bundle
has a 4-dimensional base space
and Lorentz fibers.

The Ricci tensor and scalar are given by $R_{ab} = R^{c}_{\enskip acb}$ and
$R =
\eta^{ab} R_{ab}$ and the Weyl curvature 2-form constructed from ${\bf
{R}}_{b}^{a}$
may be written as
$${\bf {C}}_{b}^{a} = {\scriptstyle {1 \over 2}} C_{\enskip bcd}^{a} {\bf
{e}}^{c}{\bf {e}}^{d} = {\bf {R}}_{b}^{a} - \Delta^{ca}_{bd} {\cal {R}}_{c}
{\bf
{e}}^{d} \eqno(2.2) $$
where
$${\cal {R}}_{a} = {\cal {R}}_{ab} {\bf {e}}^{b} \equiv -{\scriptstyle {1
\over 2}}
(R_{ab} - {\scriptstyle {1 \over 6}}\eta_{ab} R){\bf {e}}^{b} \eqno(2.3)$$
and
$$\Delta^{ca}_{bd} \equiv (\delta^{a}_{d} \delta^{c}_{b}  - \eta^{ac}
\eta_{bd})
\eqno(2.4) $$
The expression for ${\cal {R}}_{ab}$ is easily inverted to give
$$R_{ab} = -2{\cal {R}}_{ab} - \eta_{ab}{\cal {R}} \eqno(2.5) $$
Therefore, knowledge of the pair ${\bf {C}}^{a}_{b}, {\cal {R}}_{a}$ is
equivalent to
knowledge of the complete Riemann curvature tensor, ${\bf {R}}^{a}_{b}$.

Quite generally, the Bianchi identities or their generalizations may by
found by exterior
differentiation of the Maurer-Cartan structure equations.  For example,
differentiating
eq.(2.1a), then using eqs.(2.1) to replace the resulting differentials
gives the first Bianchi
identity:
$$\eqalignno{
0 \equiv {\bf {d}}^{2}{\bf {e}}^{a} &= {\bf {de}}^{b} \omega_{b}^{a} - {\bf
{e}}^{b} {\bf {d}}\omega_{b}^{a} \cr
&= {\bf {e}}^{c} \omega_{c}^{b} \omega_{b}^{a} - {\bf {e}}^{b} (\omega_{b}^{c}
\omega_{c}^{a} + {\bf {R}}_{b}^{a}) \cr
&= - {\bf {e}}^{b}{\bf {R}}_{b}^{a} &(2.6)  \cr }  $$
The cancellation of the non-curvature terms is a general property of this
type of calculation,
since it is a necessary consequence of the consistency of the original
Maurer-Cartan
structure equations for the underlying group.  Thus, we may immediately
write the second
Bianchi identity from eq.(2.1b) by replacing the differentials of the
connections with the
corresponding curvatures:
$$\eqalignno{
0 \equiv {\bf {d}}^{2} \omega_{b}^{a} &= {\bf {d}}\omega_{b}^{c}
\omega_{c}^{a}
- \omega_{b}^{c} {\bf {d}} \omega_{c}^{a} + {\bf {d}}{\bf {R}}_{b}^{a} \cr
&= {\bf {R}}_{b}^{c} \omega_{c}^{a} - \omega_{b}^{c} {\bf {R}}_{c}^{a} + {\bf
{d}}{\bf {R}}_{b}^{a}  \cr
&\equiv {\bf {DR}}_{b}^{a}  &(2.7)  \cr } $$
\medskip
We will refer to certain well-known generalizations of Riemannian
spacetimes.  One
generalization is to spaces in which the Christoffel connection acquires an
antisymmetric
part.  The antisymmetric part is a tensor field called the torsion which
may be written as a
vector-valued 2-form ${\bf {T}}^{a}$.  In a spacetime with torsion,
eqs.(2.1) are modified to
$$\eqalignno{
{\bf {de}}^{a} &= {\bf {e}}^{b} \omega_{b}^{a} + {\bf {T}}^{a} &(2.8a)  \cr
{\bf {d}}\omega_{b}^{a} &= \omega_{b}^{c} \omega_{c}^{a} + {\bf {R}}_{b}^{a}
&(2.8b) \cr}$$
and the Bianchi identity corresponding to eq.(2.8a) becomes
$${\bf {DT}}^{a} \equiv {\bf {dT}} ^{a} + {\bf {T}}^{b} \omega^{a}_{b} = {\bf
{e}}^{b}{\bf {R}}_{b}^{a} \eqno(2.9) $$

A Riemannian spacetime with trivial frame bundle is called parallelizable.
In this case there
exists a choice of the connection with torsion, for which the Riemann
curvature vanishes.

We will also use Weyl spacetimes.  Weyl spacetimes are based on the
11-dimensional Weyl group consisting of Lorentz transformations,
translations, and dilations.  The dilations may be thought of as local
changes of units (or ``gauge'' or ``scale'').  The gauge vector of these
scale changes is called the Weyl vector or Weyl form, ${\bf {W}}$.  Under
changes of scale the Weyl form changes by an exact differential.

The structure equations of a Weyl geometry are
$$\eqalignno{
{\bf {de}}^{a} &= {\bf {e}}^{b} \omega_{b}^{a} + {\bf {W}}{\bf {e}}^{b}
&(2.10a)
\cr
{\bf {d}}\omega_{b}^{a} &= \omega_{b}^{c} \omega_{c}^{a}+{\bf {R}}_{b}^{a}
&(2.10b) \cr
{\bf {d}}{\bf {W}} &= \Omega &(2.10c) \cr}$$
Eqs.(2.10) describe a connection on an 11-dim principal fiber bundle with
4-dim base space and fibers isomorphic to the homogeneous Weyl group.
The dilational curvature (or simply dilation) $\Omega$ vanishes if and only
if ${\bf
{W}}$ is closed.  A Weyl spacetime with vanishing dilation permits a choice
of scale such
that locally ${\bf {W}} = 0$.  When ${\bf {W}}$ is exact this result is
global, and the
Weyl spacetime reduces to a Riemannian spacetime.

The torsion and Weyl generalizations above may be combined to give the
structure
equations for a Weyl spacetime with torsion
$$\eqalignno{
{\bf {de}}^{a} &= {\bf {e}}^{b} \omega_{b}^{a} + {\bf {W}}{\bf {e}}^{b} + {\bf
{T}}^{a} &(2.11a) \cr
{\bf {d}}\omega_{b}^{a} &= \omega_{b}^{c} \omega_{c}^{a}+{\bf {R}}_{b}^{a}
&(2.11b) \cr
{\bf {dW}} &= \Omega &(2.11c) \cr}$$
The structure equations for biconformal geometry developed in the next
section generalize eqs.(2.11) still further.  Eqs.(2.11) are therefore a
useful guide to understanding the biconformal structure equations.

Finally, we use the simple idea of a Whitney sum bundle.  Given two vector
bundles on the same base space with fibers isomorphic to vector spaces
${\cal {V}}_1$ and ${\cal {V}}_2$, the Whitney sum bundle is that vector
bundle with the same base space and fibers isomorphic to the direct sum
vector space, ${\cal {V}}_1 \oplus {\cal {V}}_2$.
\bigskip
\noindent 3.  Biconformal geometry
\medskip

The spacetime conformal group is that 15-parameter group of transformations
which leave
the 4-dimensional Minkowski metric invariant up to an overall scale factor.
It consists of 6 Lorentz transformations, 4 translations, 4 inverse
translations\footnote{$^{1}$}{{\smallrm {The inverse translations have also
been called special conformal transformations, conformal boosts,
accelerations, or elations.  We prefer the term {\smallit {inverse
translations}} because the name better reflects their geometric
significance as translations ${\scriptstyle {( y^{\mu}\rightarrow y^{\mu} +
a^{\mu}) }}$ in inverse coordinates ${\scriptstyle {(y^{\mu} \equiv -
x^{\mu}/x^2) }}$ on compactified Minkowski space.}}}, and 1 dilation.

Following Klein [35] and Cartan [34], we construct the biconformal bundle
in two steps.  First, an elementary geometry is found by taking the
quotient, ${\cal {C}}/ {\cal {C}}_{0}$, of the conformal group, $\cal {C}$,
by its isotropy subgroup, ${\cal {C}}_{0}$. Then we generalize the
connection to that of a curved 8-dimensional space by the addition of
horizontal curvature 2-forms to the group structure equations.  This
general procedure guarantees us a Cartan connection on a principal fiber
bundle with fiber symmetry ${\cal {C}}_{0}$.

For our present purpose, what is important in this general procedure is the
identification of the isotropy subgroup, ${\cal {C}}_{0}$.  If the full
conformal group is to act effectively and transitively on the base space,
the isotropy subgroup must contain no normal subgroup of the conformal
group.  In the case of Poincar\'{e} gauging (see, eg., Utiyama [36] and
Kibble [37]) this means that the only possible principal fiber symmetry
containing the Lorentz group is the Lorentz group itself.  By contrast, the
conformal gauging leaves us more than one choice for a fiber symmetry
containing the Lorentz group.  While the standard approach to conformal
gauge theory uses the inhomogeneous Weyl group as the fiber symmetry, we
note that there is little principled distinction between the translations
and the inverse translations.  We are thus led to consider the homogeneous
Weyl group as ${\cal {C}}_{0}$, leaving all eight translations to span the
base space.  As pointed out in the introduction, this immediately solves
the auxiliary field problem, since the symmetry which was auxiliary in the
4-dimensional gauging is converted to additional coordinate freedom of the
base space in the 8-dimensional treatment.

The choice of the homogeneous Weyl group as fiber symmetry also follows if we
distinguish the conformal transformations based on their fixed points.
Noting that
translations always have exactly one fixed point when acting on {\it
{compactified}}
Minkowski space\footnote{$^{2}$}{{\smallrm {A special point and its
compactified null
cone are added at infinity to accomplish compactificaton.  See [47], [48]}}},
we find a class of eight conformal transformations (i.e., the four
translations and the four inverse translations) having a single fixed point
(infinity or the origin) while the remaining Lorentz and dilational
transformations leave both the origin and infinity fixed.

We now implement the construction, using the linear $SO(4,2)$
representation of the conformal group for our notation.  Letting the
indices $(A, B, \dots) = (0, 1, \ldots, 5)$
and $(a, b, \ldots) = (1, \ldots, 4)$, the $SO(4,2)$ metric $\eta_{AB}$
is given by $(\eta_{ab} = diag(1, 1, 1, -1), \eta_{05} = \eta_{50} = 1$ with
all other components vanishing.  Introducing the connection 1-form
$\omega_{B}^{A}$ we may express the covariant constancy of $\eta_{AB}$, as
$${\bf {D}}\eta_{AB} \equiv
{\bf {d}}\eta_{AB} - \eta_{CB} \omega_{A}^{C} - \eta_{AC} \omega_{B}^{C} = 0$$
so that $ \omega_{B}^{A} = - \eta^{AD} \eta_{BC} \omega_{D}^{C}$.  The
conformal
connection may be broken into four independent homogeneous-Weyl-invariant
parts:  the
{\it {spin connection}}, $\omega_{b}^{a}$, the {\it {solder form}},
$\omega_{0}^{a}$, the {\it {co-solder form}}, $\omega_{a}^{0}$, and the
{\it {Weyl vector}}, $\omega_{0}^{0}$ where the spin connection satisfies
$$ \omega_{b}^{a} = - \eta_{bc} \eta^{ad} \omega_{d}^{c}  \eqno (3.1a) $$
and the remaining components of $\omega_{B}^{A}$ are related by
$$ \eqalignno{
\omega_{0}^{5} &=  \omega_{5}^{0} = 0 &(3.1b) \cr
\omega_{5}^{5} &= - \omega^{0}_{0}   &(3.1c)  \cr
\omega_{5}^{a} &= - \eta^{ab} \omega_{b}^{0} &(3.1d)  \cr
\omega_{a}^{5} &= - \eta_{ab} \omega_{0}^{b}  & (3.1e) \cr } $$

\noindent   These constraints reduce the number of independent connection
forms
$\omega_{B}^{A}$  to the required 15 and allow us to restrict $(A, B,
\dots) = (0, 1,
\ldots, 4)$ in all subsequent equations (note, however, that all implicit
sums must include the equivalent ``5" index in the sum).  The structure
constants of the conformal Lie
algebra now lead immediately to the Maurer-Cartan structure equations of
the conformal
group.  These are simply
$${\bf {d\omega}}_{B}^{A} = {\bf {\omega}}_{B}^{C} {\bf
{\omega}}_{C}^{A} \eqno (3.2) $$
When broken into parts based on homogeneous Weyl transformation properties,
eq.(3.2)
gives:
$$\eqalignno{
{\bf {d\omega}}_{b}^{a} &= \omega_{b}^{c} \omega_{c}^{a} + {\bf
{\omega}}_{b}^{0} {\bf {\omega}}_{0}^{a} - \eta_{bc}\eta^{ad} {\bf
{\omega}}_{d}^{0} {\bf {\omega}}_{0}^{c} \cr
{\bf {d\omega}}_{0}^{a} &= {\bf {\omega}}_{0}^{0} {\bf {\omega}}_{0}^{a}
+ {\bf {\omega}}_{0}^{b} {\bf {\omega}}_{b}^{a} \cr
{\bf {d\omega}}_{a}^{0} &= {\bf {\omega}}_{a}^{0} {\bf {\omega}}_{0}^{0}
+  {\bf {\omega}}_{a}^{b} {\bf {\omega}}_{b}^{0} \cr
{\bf {d\omega}}_{0}^{0} &= {\bf {\omega}}_{0}^{a} {\bf {\omega}}_{a}^{0}
&  (3.3) \cr} $$

Note that the exterior derivative ${\bf {d}}$ used in eqs.(3.3) includes
partial derivatives in all eight of the base space directions.  When we use
coordinates, we will divide them into two sets with raised and lowered
index positions as in $(x^{\mu},y_{\nu})$ coresponding to the index
positions on $(\omega^{a}_{0}, \omega^{0}_{b})$.  Partial derivatives will
have correspondingly inverted positions, thus,
$$\partial_{\mu} \phi = {\partial \phi  \over  \partial x^{\mu}} \qquad
\partial^{\mu} \phi =
{\partial \phi  \over  \partial y_{\mu}}$$
These index positions provide a quick key to the scaling weights of fields,
with raised
Lorentz indices having weight $+1$ and lowered Lorentz indices weight $-1$.
No weight
is associated with $0$ indices.  In general, we do {\it {not}} freely raise
and lower indices
with a metric.

The generalization of eqs.(3.3) to a curved base space is immediate.  We have:
$$\eqalignno{
{\bf {d\omega}}_{b}^{a} &= \omega_{b}^{c}  \omega_{c}^{a} +
\omega_{b}^{0} \omega_{0}^{a} - \eta_{bc}\eta^{ad} \omega_{d}^{0}
\omega_{0}^{c} + \Omega_{b}^{a} \cr
&= \omega_{b}^{c} \omega_{c}^{a} + \Delta_{bc}^{da} \omega_{d}^{0}
\omega_{0}^{c} + \Omega_{b}^{a} &(3.4a) \cr
{\bf {d\omega}}_{0}^{a} &= \omega_{0}^{0} \omega_{0}^{a} +
\omega_{0}^{b} \omega_{b}^{a} + \Omega_{0}^{a} &(3.4b) \cr
{\bf {d\omega}}_{a}^{0} &= \omega_{a}^{0} \omega_{0}^{0} +
\omega_{a}^{b} \omega_{b}^{0} + \Omega_{a}^{0}  &(3.4c) \cr
{\bf {d\omega}}_{0}^{0} &= \omega_{0}^{a} \omega_{a}^{0} +
\Omega_{0}^{0}  &(3.4d) \cr} $$
Eqs.(3.4) are the structure equations of biconformal space.  While all of
the 2-forms $\Omega^{A}_{B}$ are curvatures, for convenience we will call
$\Omega_{b}^{a}$ the curvature.  $\Omega_{0}^{a}$ and $\Omega_{a}^{0}$ will
be called the torsion and co-torsion, respectively, and $\Omega_{0}^{0}$
the dilation.  If we set $\omega_{a}^{0}$, $\omega_{0}^{0}$ and the
corresponding co-torsion and dilation to zero, we recover the usual
structure equations for the connection of 4-dimensional spacetime with
Riemannian curvature $\Omega_{b}^{a}$ and torsion $\Omega_{0}^{a}$,
eqs.(2.8).  If we set only $\omega_{a}^{0} =  \Omega_{a}^{0} = 0$, the
structure equations are those of 4-dimensional Weyl geometry with torsion,
eqs.(2.11).

It is important to realize that the dilational curvature of a biconformal
space is a
different function of the connection than the dilational curvature of a
Weyl  geometry.
From eq.(3.4d) we see that the dilational curvature in a biconformal space is
$$ \Omega_{0}^{0} ={\bf {d\omega}}_{0}^{0} - \omega_{0}^{a} \omega_{a}^{0}$$
whereas that of a 4-dimensional Weyl geometry is given by eq.(2.10c) as
$$ \Omega  = {\bf {d}}{\bf {W}}$$
It is the dilational curvature, $\Omega$ or $\Omega_{0}^{0}$, that is
responsible for the
physical size changes discussed in the introduction.  But while the only
way to have
$\Omega = 0$ in a Weyl geometry is for the Weyl vector to be pure gauge,
${\bf {W}} =
{\bf {d}}\phi$, in biconformal space vanishing dilation, $\Omega_{0}^{0} =
0$, actually
{\it {requires}} that ${\bf {d}}\omega_{0}^{0}\neq 0$.  In [33] this fact
was used to show that biconformal space provides a satisfactory model of
the electromagnetic field, predicting the electrodynamics of a charged
point particle.  In that model, {\it {part}} of the Weyl vector is
interpreted as the electromagnetic vector potential.  In subsequent
sections of this paper we will show that another way to avoid size change
in physical models is to identify spacetime with a submanifold on which
$\Omega_{0}^{0}$ vanishes, even though $\Omega_{0}^{0} \neq 0$ for the
biconformal space as a whole.

A second important  fact about the 8-dimensional gauging of the conformal
group is seen from the effect of gauge transformations.  For the full
conformal group, eqs.(3.3), $\omega_{B}^{A}$ transforms according to
$${\hat{\omega}}_{B}^{A} = \Lambda^{A}_{\enskip C} \> \omega_{D}^{C}\>
{\bar{\Lambda}}^{D}_{\enskip B} - {\bf {d}}\Lambda^{A}_{\enskip C}
{\bar{\Lambda}}^{C}_{\enskip B} \eqno(3.5) $$
where ${\bar{\Lambda}}^{A}_{\enskip B}$ is the inverse of the local SO(4,2)
transformation $\Lambda^{A}_{\enskip B}$.  Relations analagous to eqs.(3.1)
hold between the different components of $\Lambda^{A}_{\enskip B}$, and
similarly allow the restriction $A, B \ldots = 0, 1, \ldots , 4$ (except
for sums).  Written in terms of the Lorentz covariant parts eq.(3.5)
becomes, for example,
$$\eqalignno{
{\hat{\omega}}_{b}^{a} =&
\Lambda^{a}_{\enskip c} \> \omega_{d}^{c}\> {\bar{\Lambda}}^{d}_{\enskip b} +
\Lambda^{a}_{\enskip c} \> \omega_{0}^{c}\> {\bar{\Lambda}}^{0}_{\enskip b}
-  \eta^{ac} \eta_{bf} \Lambda^{d}_{\enskip c} \>  \omega_{d}^{0}\>
{\bar{\Lambda}}^{f}_{\enskip 0} + \Lambda^{a}_{\enskip 0} \>
\omega_{d}^{0}\> {\bar{\Lambda}}^{d}_{\enskip b} + \Lambda^{a}_{\enskip 0}
\> \omega_{0}^{0}\> {\bar{\Lambda}}^{0}_{\enskip b} \cr
 &- \eta^{ac} \eta_{bd} \Lambda^{0}_{\enskip c} \> \omega_{0}^{f}\>
{\bar{\Lambda}}^{d}_{\enskip f} - \eta^{ac} \eta_{bd} \Lambda^{0}_{\enskip
c} \> \omega_{0}^{0}\> {\bar{\Lambda}}^{d}_{\enskip 0} - {\bf
{d}}\Lambda^{a}_{\enskip c} {\bar{\Lambda}}^{c}_{\enskip b} - {\bf
{d}}\Lambda^{a}_{\enskip 0} {\bar{\Lambda}}^{0}_{\enskip b} - \eta^{ac}
\eta_{bd} {\bf {d}}\Lambda^{0}_{\enskip c} {\bar{\Lambda}}^{d}_{\enskip 0}
&(3.6) \cr } $$
Similar expressions hold with the indices $a$ and/or $b$ replaced by $0$.
Now, when the gauging is over a 4-dimensional base manifold, the bundle no
longer posesses translational invariance, so all terms containing
$\Lambda^{a}_{\enskip 0}$ drop out, leaving
$$\eqalignno{
{\hat{\omega}}_{b}^{a} &=
\Lambda^{a}_{\enskip c} \> \omega_{d}^{c}\> {\bar{\Lambda}}^{d}_{\enskip b}
+ \Lambda^{a}_{\enskip c} \> \omega_{0}^{c}\> {\bar{\Lambda}}^{0}_{\enskip b}
- \eta^{ac} \eta_{bd} \Lambda^{0}_{\enskip c} \> \omega_{0}^{f}\>
{\bar{\Lambda}}^{d}_{\enskip f} - {\bf {d}}\Lambda^{a}_{\enskip c}
{\bar{\Lambda}}^{c}_{\enskip b} &(3.7a) \cr
{\hat{\omega}}_{0}^{a} &=
\Lambda^{a}_{\enskip b} \> \omega_{0}^{b}\> {\bar{\Lambda}}^{0}_{\enskip 0}
&(3.7b)  \cr
{\hat{\omega}}_{a}^{0} &=
\Lambda^{0}_{\enskip c} \> \omega_{d}^{c}\> {\bar{\Lambda}}^{d}_{\enskip a}
+ \Lambda^{0}_{\enskip b} \> \omega_{0}^{b}\> {\bar{\Lambda}}^{0}_{\enskip a}
+  \Lambda^{0}_{\enskip 0} \> \omega_{b}^{0}\> {\bar{\Lambda}}^{b}_{\enskip a}
+  \Lambda^{0}_{\enskip 0} \> \omega_{0}^{0}\> {\bar{\Lambda}}^{0}_{\enskip a}
- {\bf {d}}\Lambda^{0}_{\enskip b} {\bar{\Lambda}}^{b}_{\enskip a}
- {\bf {d}}\Lambda^{0}_{\enskip 0} {\bar{\Lambda}}^{0}_{\enskip a}  &(3.7c) \cr
{\hat{\omega}}_{0}^{0} &=  \omega_{0}^{0}
+ \Lambda^{0}_{\enskip b} \> \omega_{0}^{b}\> {\bar{\Lambda}}^{0}_{\enskip 0}
- {\bf {d}}\Lambda^{0}_{\enskip 0} {\bar{\Lambda}}^{0}_{\enskip 0} &(3.7d)
\cr } $$
Notice that the solder form $\omega_{0}^{a}$ becomes tensorial with respect
to the fiber symmetry.  This is what makes the spacetime metric tensorial
when general relativity is treated as a Poincar\'{e} gauge theory.

In biconformal space, the base manifold is spanned by both $\omega_{0}^{a}$
and $\omega_{a}^{0}$ together.  This removes both translations from the
fiber symmetry, leaving
$$\eqalignno{
{\hat{\omega}}_{b}^{a} &=
\Lambda^{a}_{\enskip c} \> \omega_{d}^{c}\> {\bar{\Lambda}}^{d}_{\enskip b}
- {\bf {d}}\Lambda^{a}_{\enskip c} {\bar{\Lambda}}^{c}_{\enskip b} &(3.8a) \cr
{\hat{\omega}}_{0}^{a} &= \Lambda^{a}_{\enskip b} \> \omega_{0}^{b}\>
{\bar{\Lambda}}^{0}_{\enskip 0}  &(3.8b) \cr
{\hat{\omega}}_{a}^{0} &=  \Lambda^{0}_{\enskip 0} \> \omega_{b}^{0}\>
{\bar{\Lambda}}^{b}_{\enskip a} &(3.8c) \cr
{\hat{\omega}}_{0}^{0} &=  \omega_{0}^{0}
- {\bf {d}}\Lambda^{0}_{\enskip 0} {\bar{\Lambda}}^{0}_{\enskip 0} &(3.8d)
\cr } $$
Eqs.(3.8) are the gauge transformations of the biconformal connection.  Now
both sets of solder forms are tensorial.  Also, even though the base space
is 8-dimensional, the Lorentz transformations remain 4-dimensional matrices
(but with 8-dimensional functional dependence), with the co-space indices
transforming with inverse Lorentz transformations.  As a result, there are
many tensorial components to the curvature.  Indeed, not only are the
curvature, torsion, co-torsion and dilation tensorial, transforming as
$$\eqalignno{
{\hat{\Omega}}_{b}^{a} &= \Lambda_{\enskip c}^{a} \Omega_{d}^{c}
{\bar{\Lambda}}_{\enskip b}^{d} &(3.9a) \cr
{\hat{\Omega}}_{0}^{a} &= \Lambda_{\enskip b}^{a} \Omega_{0}^{b}
{\bar{\Lambda}}_{\enskip 0}^{0} &(3.9b) \cr
{\hat{\Omega}}_{a}^{0} &= \Lambda_{\enskip 0}^{0} \Omega_{b}^{0}
{\bar{\Lambda}}_{\enskip a}^{b} &(3.9c) \cr
{\hat{\Omega}}_{0}^{0} &=  \Omega_{0}^{0} &(3.9d) \cr }$$
but it is important to remember that each of these separate curvatures
takes the general form
$$\Omega_{B}^{A} = {\scriptstyle {1 \over 2}} \> \Omega_{Bcd}^{A} \>
\omega_{0}^{c}  \omega_{0}^{d} + \Omega_{Bd}^{Ac} \> \omega_{c}^{0}
 \omega_{0}^{d} + {\scriptstyle {1 \over 2}} \> \Omega_{B}^{Acd}\>
\omega_{c}^{0} \omega_{d}^{0} \eqno (3.10) $$
Recalling that the solder and co-solder forms are tensorial, the three
terms of these curvatures will not mix under the Lorentz transformations
and dilations of the fiber group.
Based on the relationship between biconformal space and one-particle phase
space demonstrated in [33], and on the results of Sec.(6) we will call
$\Omega^{A}_{Bcd} = \Omega^{A}_{B[cd]}$ the spacetime term,
$\Omega^{Ac}_{Bd}$ the cross term, and
$\Omega^{Acd}_{B} = \Omega^{A[cd]}_{B}$ the momentum term of each
homogeneous-Weyl-invariant type of curvature.  Notice that while these
terms do not mix
along the fibers of the biconformal bundle, they do mix under changes of
the basis
$(\omega_{0}^{a}, \omega_{a}^{0})$.  Finally we note that the 2-form, ${\bf
{d}}\omega_{0}^{0}$, is also invariant.
\medskip
We now turn to the metric structure of biconformal space.  While the
$SO(4,2)$ metric is preserved by conformal transformations, this
6-dimensional metric cannot be straightforwardly used to give a metric on
the 8-dimensional base space.  However, every biconformal space has another
natural metric structure based on the Killing metric.  The Killing metric
is built from the conformal group structure constants as
$$K_{AB} = {\scriptstyle {- {1 \over 16}}} C^{C}_{AD} C^{D}_{BC} = \pmatrix{
{\scriptstyle {- {1 \over 2}}}\Delta_{db}^{ac} & 0 & 0 & 0 \cr
0 & 0 & \delta^{b}_{a} & 0 \cr
0 &  \delta^{a}_{b}& 0 & 0 \cr
0 & 0 & 0 & {\scriptstyle {- {1 \over 2}}} \cr}  \eqno(3.11)$$
where
$$(A, B, \ldots) \in \{ {\scriptstyle {{0\choose 0} , {a\choose 0} ,
{0\choose a} ,
{a\choose b}}} \}.$$
When this metric is restricted to the base manifold, $A$ and $B$ are
restricted to $\{
{a\choose 0} , {0\choose a}  \}$ and the metric takes the form
$$\eqalignno{
K_{AB} &= \pmatrix{
K_{ab} & K^{a}_{b} \cr
K^{b}_{a} & K^{ab} \cr} \cr
&= \pmatrix{
0 & \delta^{a}_{b} \cr
\delta^{b}_{a} & 0 \cr}  &(3.12) \cr}$$
This is clearly invariant under homogeneous Weyl transformations.  The
metric $K_{AB}$
has all eigenvalues equal to $\pm1$ and zero signature, so that given the
Lorentz structure
of the underlying 4-dimensional geometry the diagonal form is necessarily
$diag(\eta_{ab}, - \eta^{ab})$.
The metric provides an indefinite inner product between the basis forms,
$\langle
\omega^{A}, \omega^{B}
\rangle = K^{AB}$, or
$$\eqalignno{
\langle \omega^{a}_{0}, \omega_{0}^{b} \rangle &= 0 \cr
\langle \omega^{a}_{0}, \omega^{0}_{b} \rangle &= \delta^{a}_{b} \cr
\langle \omega^{0}_{a}, \omega_{b}^{0} \rangle &= 0 &(3.13) \cr }$$
where $K^{AB} \equiv (K^{-1})_{AB} = K_{AB}$.  Expanding in a coordinate
basis we write
$$\omega^{A} = (\omega^{a}_{0}, \omega^{0}_{a}) = \omega_{\enskip M}^{A} {\bf
{d}}x^{M}  \eqno(3.14) $$
where $M \in \{  {\mu\choose} , { \choose \nu}  \}$ and $x^{M} =
(x^{\mu}, y_{\nu}).$  The components of the metric become
$$\eqalignno{
g^{MN} &= \langle {\bf {d}}x^{M}, {\bf {d}}x^{N} \rangle  \cr
&=K^{AB} \omega^{\enskip M}_{A} \omega^{\enskip N}_{B}  &(3.15) \cr }$$
with $ g^{MN}$ inverse to $g_{MN}$ in the usual way.
Notice that the metric structure in this conformal gauge theory is
intrinsic to the base space while the metric structure of Poincar\'{e}
gauge theory relies on a coincidental identification between two spaces.
To see this difference clearly, recall that gauge theory makes use of {\it
{two}} different (possibly metric) spaces:  (1) the (gauged or ungauged)
group manifold itself, the base space of which becomes the physical space
and (2) the representation space on which the group transformations act.
These spaces are not {\it {a priori}} related.  In Poincar\'{e} gauging the
projection of the Poincar\'{e} Killing metric onto the 4-dimensional base
space of the group manifold vanishes.  The existence of the usual spacetime
metric used when general relativity is regarded as a Poincar\'{e} gauge
theory is therefore {\it {not}} an intrinsic property of the group
manifold, but instead occurs because the the representation space is
coincidentally isomorphic to the base space of the ungauged group manifold.
The same metric may therefore be used.  By contrast, for conformal gauging
the projected Killing metric is nondegenerate, and is already defined on
the group manifold.  The coincidence which was required by the Poincar\'{e}
gauging does not even occur for the conformal group, where the
representation metric $\eta_{AB}$ is 6-dimensional.  But the conformal
group does not require a coincidence.

The origin of the metric in gauging the conformal group also gives an
additional motivation for choosing the 8-dimensional gauging used here over
the 4-dimensional conformal gauging used elsewhere [38 - 44].  Notice that
when $K_{AB}$ is projected onto only a 4-dimensional base space, it again
vanishes.  Thus, 4-dimensional conformal gauge theory also relies on a
coincidental isomorphism for its metric structure.

While this metric structure is important in biconformal spaces, it is
necessary to realize that
not all orthonormal frames may be reached by motion on the fibers, which
provide only {\it
{4-dim}} Lorentz transformations and dilations.  Thus, the 15-dimensional
biconformal bundle is much more rigidly structured than the full
36-dimensional bundle of orthonormal frames of an 8-dimensional
pseudo-Riemannian geometry.  It is this restricted fiber symmetry that
allows the invariance of both the $SO(4,2)$ metric and the Killing metric.
\bigskip
With our underlying structures thus established, our goal now is to find a
1-1 relationship between 4-dimensional pseudo-Riemannian spacetimes and a
subclass of biconformal spaces, and in the process to gain insights into
what new features are possessed by the biconformal spaces.  We are
especially interested in formulating general relativity within the context
of conformal gauging, anticipating that the extra conformal fields may
supply some insight into the known fundamental structure beyond general
relativity.  The main result of this paper accomplishes the desired 1-1
mapping.

Our central theorem states a set of necessary and sufficient
biconformally-invariant
conditions for the biconformal bundle to be homeomorphic to a Riemannian
spacetime.  This equivalence requires both the existence of a
14-dimensional $\omega_{0}^{0} = 0$ sub-bundle of the biconformal bundle,
and an isomorphism between four dimensions of the biconformal base manifold
and the tangent space fibering over the Riemannian spacetime.  The
homeomorphism therefore identifies the 14-dimensional biconformal
sub-bundle with the Whitney sum of the Riemannian bundle of orthonormal
frames and the Riemannian tangent bundle.  The resulting dual character of
a normal biconformal space as simultaneously a 4-dimensional Riemannian and
an 8-dimensional biconformal space allows us to easily write the field
equation of general relativity in a biconformally invariant manner.

The central theorem may also be expressed in terms of the full
15-dimensional biconormal
bundle, in which case the homeomorphism is with the triple Whitney sum of
the bundle of orthonormal frames, the tangent bundle, and the bundle of
Weyl gauges.

In addition to the proof of the main theorem, we show a class of
biconformal spaces in which there is a closed 2-form, independent of the
Weyl vector, which can be interpreted as an arbitrary electromagnetic
field.  In concluding we present biconformal field equations including
phenomenological stress-energy tensor and electromagnetic current density
which yield the full Einstein and Maxwell equations, respectively.

The necessary and sufficient conditions of the proof (described further in
subsequent sections) are the following:
\parindent=.75in
\item{1.} Vanishing torsion, $\Omega^{a}_{0} = 0$.
\item{2.} Symplectic dilation:  The dilation $\Omega_{0}^{0}$ is closed and
the Weyl
vector $\omega_{0}^{0}$ is exact.
\item{3.} Trace-free spacetime curvature:  $\Omega^{a}_{bac} = 0$.
\medskip
\parindent=.25in
\noindent Conditions (2) and (3) require some clarification.  Condition (2)
calls the dilation symplectic because when $\omega_{0}^{0}$ is exact the
dilational structure equation requires $\Omega_{0}^{0}$ to be
nondegenerate.  Being both closed and nondegenerate, $\Omega_{0}^{0}$ is
symplectic [50].  Condition (3) holds only in a particular class of frames.
The existence of this class of frames is guaranteed by conditions (1 \& 2).
A biconformal space satisfying conditions (1-3) will be called {\it
{normal}}.

We note in passing that the symplectic condition (2) gives the biconformal
manifold both symplectic and almost complex structure.  It is a simple
matter to rewrite the biconformal structure equations as a real homogeneous
Weyl fiber bundle on a 4-dimensional complex base manifold.  These
implications of these structures will be examined elsewhere.
\bigskip
\noindent 4.  Torsion-free biconformal spaces
\medskip
Our goal is to find necessary and sufficient conditions for the existence
of a 1-1 relationship between 4-dimensional pseudo-Riemannian spacetimes
and a well-defined subclass of biconformal spaces.  In this section we find
the consequences for the biconformal connection of vanishing torsion, the
first of the necessary and sufficient conditions.  Vanishing torsion
guarantees that the solder form is in involution, allowing the use of the
Frobenius theorem to constrain the form of the connection.

Two remarks about the vanishing torsion condition are in order.  First, we
note that an
involution for $\omega^{a}_{0}$ requires vanishing of only the momentum
term of the
torsion.  However, since consistency with general relativity requires
vanishing spacetime
torsion, it is natural to begin by assuming the full torsion tensor is
zero.  The included
vanishing of the cross-term is not a severe restriction, since there remain
numerous
unspecified fields among the curvature, the co-torsion and the dilation.

Second, we will allow in this section the possibility of a separate
assumption that the foliation provided by the solder form involution is
regular.  A foliation of a space ${\cal {S}}$ is {\it {regular}} if
$\forall$ points $P, \exists$ a neighborhood ${\cal {N}}_{P} \subset {\cal
{S}}$ such that the leaf on which $P$ lies intersects ${\cal {N}}_{P}$ only
once. Regularity is the necessary and sufficient condition for the set of
orbits to be a manifold. Thus, demanding regularity of the solder form
foliation insures that the space of leaves is a 4-dimensional manifold.
This 4-dimensional manifold is identified with spacetime.  This is a useful
assumption if one wishes to study torsion-free biconformal spaces without
the additional assumptions of normality.  For normal biconformal spaces,
when all of conditions (1-3) hold, the regularity assumption is unnecessary
because a second involution guarantees the existence of the spacetime
sub-manifold.  This automatic regularity in normal biconformal spaces is
actually an improvement on the constuction of general relativity from
Poincar\'{e} symmetry principles, for which the involution of the vierbein
must be assumed regular.
\medskip
\parindent=.5in
\hang
\noindent{\it {Def:}}  A {\it {regular torsion-free biconformal space}} is
a biconformal
space with $\Omega^{a}_{0} = 0$ and such that the resulting foliation is
regular.
\medskip
\parindent=.25in
As stated above, the torsion-free condition $\Omega_{0}^{a} = 0$ places the
solder form of a biconformal space in involution.  As a result, the
Frobenius theorem guarantees that the biconformal bundle is foliated by
11-dimensional manifolds, each of which is a principal sub-bundle with a
4-dimensional base manifold and 7-dimensional homogeneous Weyl group
fibers.  With these observations we begin our lemma.
\medskip
\parindent=.5in
\hang
\noindent{\it {Lemma:}}  The connection of a (regular) torsion-free
biconformal space may
(locally) be put in the form
$$\eqalign{
\omega_{b}^{a} &= C_{\enskip bc}^{a} {\bf {e}}^{c}  \cr
\omega_{0}^{a} &= {\bf {e}}^{a} \cr
\omega_{a}^{0} &= {\bf {f}}_{a} +  B_{ab} {\bf {e}}^{b}  \cr
\omega_{0}^{0} &= W_{a} {\bf {e}}^{a}  \cr}$$
where ${\bf {e}}^{a} = e_{\mu}^{\enskip a}(x){\bf {d}}x^{\mu}$, ${\bf
{f}}_{a} = f^{\enskip \mu}_{a}(x,y) {\bf {d}}y_{\mu}$ and the coefficients
$W_{a}, B_{ab}$ and
$C_{\enskip bc}^{a}$ are functions of $x^{\mu}$ and $y_{\mu}$.  The
$y_{\mu}$ are four independent coordinates spanning the base manifold of
the sub-bundle, while the $x^{\mu}$-coordinates are either coordinates on
the manifold of orbits (when we assume the foliation regular), or are
simply four coordinates independent of the $y$-coordinates as guaranteed by
the definition of a foliation.
\medskip

\parindent=.5in

\noindent{\it {Proof:}} \enskip  Imposing vanishing torsion,
$\Omega_{0}^{a} = 0$, the
structure equations take the form

\leftskip=.5in
\rightskip=.25in
\parindent=.25in
$$\eqalignno{
{\bf {d}}\omega_{b}^{a} &= \omega_{b}^{c} \omega_{c}^{a} + \Delta_{bc}^{da}
\omega_{d}^{0} \omega_{0}^{c} + \Omega_{b}^{a} &(4.1a)\qquad \cr
{\bf {d}}\omega_{0}^{a} &= \omega_{0}^{0} \omega_{0}^{a} +  \omega_{0}^{b}
\omega_{b}^{a}  & (4.1b)\qquad \cr
{\bf {d}}\omega_{a}^{0} &= \omega_{a}^{0} \omega_{0}^{0} + \omega_{a}^{b}
\omega_{b}^{0}+ \Omega_{a}^{0} & (4.1c)\qquad \cr
{\bf {d}}\omega_{0}^{0} &= \omega_{0}^{a} \omega_{a}^{0}+ \Omega_{0}^{0} &
(4.1d) \qquad \cr}$$
The first step is to make use of the Bianchi identity following from
eq.(4.1b).  As
discussed in Sec.(2), we take the exterior derivative, ${\bf {d}}^{2}
\omega_{0}^{a} =
0$ and replace all resulting differentials of connection forms on the right
side of the
equation using the corresponding structure equations.  This leads to
$$\omega^{b}_{0} \Omega^{a}_{b} = \Omega_{0}^{0} \omega^{a}_{0} \eqno(4.2)
\qquad $$
The momentum term of this expression requires
$${\scriptstyle {{1  \over 2}}} \omega^{b}_{0} \Omega^{acd}_{b}\omega_{c}^{0}
\omega_{d}^{0} = {\scriptstyle {{1  \over 2}}}\delta_{b}^{a} \Omega_{0}^{0cd}
\omega^{b}_{0} \omega_{c}^{0} \omega_{d}^{0} \eqno(4.3) \qquad $$
from which it immediately follows that
$$\Omega^{acd}_{b} = \delta_{b}^{a} \Omega_{0}^{0cd} \eqno(4.4) \qquad $$
and since $\eta_{ae}\Omega^{ecd}_{b} = - \eta_{be}\Omega^{ecd}_{a}$ we have
$\Omega^{acd}_{b} = 0 = \Omega_{0}^{0cd}$.  We will not need the additional
constraints from the spacetime- and cross-terms of eq.(4.2) and no further
new constraints follow from the remaining Bianchi identities.

The system may now be further simplified by making use of the involution of
eq.(4.1b).
This allows us to consistently set $\omega_{0}^{a} = 0$ and first study the
sub-bundle
spanned by the remaining eleven 1-forms.  By the Frobenius theorem there
exist coordinates $x^{\mu}$ such that each constant value of $x^{\mu}$
singles out one leaf of the foliation.  Thus, the solder form
$\omega_{0}^{a} = 0$ must be of the form $\omega_{0}^{a} = {\bf {e}}^{a} =
e_{\mu}^{\enskip a}(x, y) {\bf {d}}x^{\mu}$, where $y_{\mu}$ are four
additional independent coordinates.  Also, let
 $${\bf {f}}_{a} \equiv f_{a}^{\enskip \mu}(x, y) {\bf {d}}y_{\mu} \equiv
\omega^{0}_{a} \vert_{x=const.}$$

Each 11-dimensional $x^{\mu} = const.$ submanifold has the reduced set of
structure equations
$$\eqalignno{
{\bf {d}}\omega_{b}^{a} &= \omega_{b}^{c} \omega_{c}^{a}  & (4.5a) \qquad \cr
{\bf {df}}_{a} &= {\bf {f}}_{a} \omega_{0}^{0} + \omega_{a}^{b} {\bf
{f}}_{b} +
\Omega^{0}_{a} &(4.5b) \qquad \cr
{\bf {d}}\omega_{0}^{0} &= 0 & (4.5c) \qquad \cr}$$
which may be recognized as those of a 4-dimensional Weyl geometry with zero
Riemannian curvature, zero dilation and nonvanishing
torsion\footnote{$^{3}$}{{\smallrm {Vanishing curvature implies
parallelizability of the leaves.  Therefore, unless one performs a
perversely irrational pasting together of opposite sides of the momentum
base space, the regularity of the foliation is automatic.}}}.

Since eq.(4.5c) implies a pure-gauge form for the Weyl vector we choose the
gauge so that
the Weyl vector $\omega_{0}^{0}$ vanishes locally on each subspace, leaving
the structure equations for a parallelizable Riemannian geometry.
Similarly, the involution for $\omega^{a}_{b}$ means that we can find a
local Lorentz transformation making the spin connection vanish as well, and
we are left with a connection on the reduced bundle of the form
$$\eqalignno{
\omega_{b}^{a} &= \omega_{0}^{0} = 0  & (4.6a) \qquad \cr
{\bf {df}}_{a} &=  \Omega^{0}_{a} &(4.6b)\qquad \cr }$$

Next, we reintroduce the remaining four independent 1-forms by allowing
$x^{\mu}$ to
vary.  From the linearity of the basis 1-forms in ${\bf {d}}x^{a}$,
$\omega_{b}^{a}$,
$\omega_{a}^{0}$ and $\omega_{0}^{0}$ will change only by terms
proportional to
${\bf {e}}^{a}$.  This gives an extension from a neighborhood in the
11-dimensional sub-bundle to a neighborhood in the full 15-dimensional
bundle on which the connection takes the form:
$$\eqalignno{
\omega_{b}^{a} &= C_{\enskip bc}^{a} {\bf {e}}^{c}  & (4.7a) \qquad \cr
\omega_{0}^{a} &= {\bf {e}}^{a} & (4.7b) \qquad \cr
\omega_{a}^{0} &= {\bf {f}}_{a} +  B_{ab} {\bf {e}}^{b} & (4.7c) \qquad \cr
\omega_{0}^{0} &= W_{a} {\bf {e}}^{a} & (4.7d) \qquad \cr}$$
where the coefficients $W_{a}, B_{ab}$ and $C_{\enskip bc}^{a}$ are
functions of
$x^{a}$ and $y_{a}$.

Finally, consider eq.(4.1b) with the connection in the form above.
$$\eqalignno{
{\bf {de}}^{a} &= \partial_{\mu} e_{\nu}^{a} {\bf {d}}x^{\mu}{\bf
{d}}x^{\nu} +
\partial^{\mu} e_{\nu}^{a} {\bf {d}}y_{\mu}{\bf {d}}x^{\nu} \cr
&= W_{b} {\bf {e}}^{b} {\bf {e}}^{a} +  {\bf {e}}^{b} C_{\enskip bc}^{a}{\bf
{e}}^{c}  & (4.1b) \qquad \cr } $$
Since the solder form is proportional to ${\bf {d}}x^{\mu}$ but not ${\bf
{d}}y_{\mu}$,
the presence of only a single ${\bf {d}}y_{\mu} {\bf {d}}x^{\nu}$
cross-term requires
$\partial^{\mu} e_{\nu}^{a} = 0$.  Therefore, ${\bf {e}}^{a}$ depends only on
$x^{\mu}$, completing the lemma.
\medskip

\leftskip=0in
\rightskip=0in
\parindent=.25in
The form of the co-solder form found in eq.(4.7c) suggests the importance
of the class of
co-solder forms ${\bf {f}}_{a}$, which, together with ${\bf {e}}^{a}$, span
the base
space.  Clearly, different choices of $y$-coordinate lead to different
${\bf {f}}_{a}$ and
$B_{ab}$, with $B_{ab}$ transforming inhomogeneously.  We end this section
with the
definition:
\medskip
\parindent=.5in
\hang
\noindent{\it {Def:}}  An {\bf {e}}{\it {-co-basis}} is any collection of
four 1-forms
which together with ${\bf {e}}^{a}$, span the base space of a regular,
torsion-free
biconformal space.
\medskip

\parindent=.25in
\bigskip
\noindent 5.  Normal biconformal spaces
\medskip

We now turn to our central theorem, showing that the conditions at the end
of Sec.(3) are necessary and sufficient for a biconformal space to reduce
to a vector bundle with a Riemannian spacetime as the base manifold.

A few comments may make the usefulness of the idea of a ``normal" space
more transparent.  It is helpful to compare the concept of a normal
biconformal space to a corresponding concept for general differential
geometries with metric, with Riemannian geometry playing the role of a
``normal" differential metric geometry.  A general differential geometry
with metric consists of a manifold with a metric and a connection. In
general, the connection will be neither symmetric, nor compatible with the
metric.  Vanishing torsion and vanishing covariant derivative of the metric
(non-metricity) provide necessary and sufficient conditions for the
differential geometry to be Riemannian.  What has been achieved, in fact,
is uniqueness.  These constraints on the geometry allow the metric (on a
given manifold) to uniquely fix the connection.  At the same time, the
definition of a Riemannian geometry in terms of vanishing torsion and
non-metricity tensors gives a classification of non-Riemannian geometries.
Each invariant part (using symmetries, traces, etc.) of the torsion or
non-metricity which is taken nonzero provides a distinct class of
generalizations of Riemannian geometry.

Similarly, our definition of a normal biconformal space simultaneously
provides a rule for
deriving a unique biconformal space from a 4-dimensional spacetime metric,
and implicitly provides a classification of non-normal biconformal spaces.
In this section we prove the uniqueness, and in Sec.(6) we make use of a
simple non-normal space to propose a new geometric theory of large-scale
electromagnetism.
\medskip
\parindent=.5in
\hang
\noindent{\it {Def:}}  A {\it {normal biconformal space}} is a torsion-free
biconformal
space with exact Weyl form and closed dilational curvature, for which there
exists an ${\bf
{e}}$-co-basis such that $\Omega^{c}_{acb} = 0$.
\medskip
\noindent  It turns out that the additional conditions of the definition
allow us to drop the
regularity requirement from the torsion constraint.  We now state our
central theorem.
\medskip
\parindent=.7in
\hang
\noindent{\it {Theorem:}}  The $\omega_{0}^{0} = 0$ sub-bundle of a normal
biconformal space, with fiber gauge transformations restricted to
$\Lambda^{a}_{\enskip b} = \Lambda^{a}_{\enskip b}(x)$ and
$\Lambda^{0}_{\enskip 0} = const.$ is homeomorphic to the Whitney sum
bundle of the tangent bundle and the bundle of orthonormal Lorentz frames
over a 4-dimensional pseudo-Riemannian spacetime with solder form ${\bf
{e}}^{a}$.
\medskip

\parindent=.5in
\rightskip=.5in
\hang
\noindent{\it {Proof:}} First, because $\omega_{0}^{0}$ is exact we can
perform a local
rescaling to remove it globally.  This places all that follows on a
14-dimensional sub-bundle of the biconformal bundle.  Now by lemma 1,
noting that having $\omega_{0}^{0} = 0$ from the start does not alter the
proof, the connection may be written in the form:
$$\eqalignno{
\omega_{b}^{a} &= C_{\enskip bc}^{a} {\bf {e}}^{c} \hskip .5in & (4.7a)
\hskip .5in \cr
\omega_{0}^{a} &= {\bf {e}}^{a} \hskip .5in &(4.7b) \hskip .5in \cr
\omega_{a}^{0} &= {\bf {f}}_{a} +  B_{ab} {\bf {e}}^{b} \hskip .5in &
(4.7c) \hskip .5in \cr
\omega_{0}^{0} &= 0 \hskip .5in &(4.7d') \hskip .5in \cr}$$
where ${\bf {f}}_{a}$ is chosen as the (or any) ${\bf {e}}$-co-basis such that
$\Omega^{c}_{acb} = 0$.  Then eq.(4.1b) becomes
$${\bf {de}}^{a} =  {\bf {e}}^{b} \omega_{b}^{a} \hskip .5in \eqno(5.1)
\hskip .5in $$
In the usual way, eq.(5.1) may be solved uniquely for the spin connection
in terms of the
solder form, its inverse, and its first derivatives, making the spin
connection purely $x$-dependent.
$$\omega_{b}^{a} = \omega_{b}^{a}({\bf {e}}^{b}(x)) \hskip .5in  \eqno(5.2)
\hskip .5in $$
Clearly, the vanishing of $\omega_{0}^{0}$ and the $x$-dependence of
$\omega_{b}^{a}$ will be preserved iff we restrict any further gauge
transformations to the form $\Lambda^{0}_{\enskip 0} = const.$ for scalings
and $\Lambda^{a}_{\enskip b} = \Lambda^{a}_{\enskip b}(x)$ for Lorentz
transformations.

\leftskip=.5in
\parindent=.25in
Now consider the curvature, given by eq.(4.1a).  Expanding $\Omega_{b}^{a}$
in the $({\bf {e}}^{a}, {\bf {f}}_{a})$ basis, and noting that the
vanishing of the momentum term $\Omega_{b}^{acd}$ in the $(\omega_{0}^{a},
\omega_{a}^{0})$ basis guarantees its vanishing in the $({\bf {e}}^{a},
{\bf {f}}_{a})$ basis, we have
$${\bf {d}}\omega_{b}^{a} = \omega_{b}^{c} \omega_{c}^{a} + \Delta_{bc}^{da}
({\bf {f}}_{d} +  B_{dh} {\bf {e}}^{h}) {\bf {e}}^{c} + {\scriptstyle {{1
\over 2}}}
\Omega_{bcd}^{a}{\bf {e}}^{c}{\bf {e}}^{d} + \Omega_{bd}^{ac}{\bf {f}}_{c}{\bf
{e}}^{d}  \eqno(5.3) \hskip .5in  $$
Since we may identify the expression ${\bf {d}}\omega_{b}^{a} - \omega_{b}^{c}
\omega_{c}^{a}$ with the usual 4-dimensional Riemannian curvature 2-form,
the spacetime and cross-term parts of eq.(5.3) are respectively
$$\eqalignno{
\Omega_{bcd}^{a}{\bf {e}}^{c}{\bf {e}}^{d} &= {\bf {R}}_{\enskip bcd}^{a}{\bf
{e}}^{c}{\bf {e}}^{d} - 2\Delta_{bd}^{ha} B_{hc} {\bf {e}}^{c} {\bf
{e}}^{d} \hskip .5in &(5.4) \hskip .5in  \cr
\Omega_{bd}^{ac}{\bf {f}}_{c}{\bf {e}}^{d} &= \Delta_{bd}^{ca} {\bf
{f}}_{c} {\bf
{e}}^{d} \hskip .5in &(5.5) \hskip .5in  \cr} $$
Now the trace-free condition $\Omega_{acb}^{c} = 0$ holds if and only if
the expression on the right hand side of eq.(5.4) is the trace-free part of
the Riemann curvature, namely, the Weyl curvature.  Eq.(2.2) for the Weyl
curvature then lets us immediately identify
$$B_{ab}= {\cal {R}}_{ab} \hskip .5in \eqno(5.6) \hskip .5in  $$
and eq.(5.4) becomes ${1 \over 2}\Omega_{bcd}^{a}{\bf {e}}^{c}{\bf {e}}^{d}
= {\bf {C}}_{b}^{a}$.

Now, since $B_{ab}= {\cal {R}}_{ab}$ is symmetric, eq.(4.1d) reduces to
$$\Omega_{0}^{0} = {\bf {f}}_{a}{\bf {e}}^{a} \hskip .5in \eqno(5.7) \hskip
.5in  $$
We note in passing that eq.(5.7) shows that $\Omega_{0}^{0}$ is necessarily
nondegenerate.  Because it is assumed closed, it is symplectic [50].  The
Bianchi identity for eq.(4.1d) reduces to
$${\bf {e}}^{a}\Omega_{a}^{0}=0 \eqno(5.8) \hskip .5in $$
which implies vanishing momentum term for the co-torsion, $\Omega_{a}^{0bc}
= 0$.
Therefore, on the $x=const.$ sub-manifolds, ${\bf {f}}_{a}$ satisfies
$${\bf {df}}_{a}\vert_{x=const.}=0 \eqno(5.9) \hskip .5in $$
and we can find coordinates $y_{\mu}$ such that
$${\bf {f}}_{a} = f_{a}^{\enskip \mu}(x){\bf {d}}y_{\mu}$$
Thus, $\Omega_{0}^{0} =  f_{a}^{\enskip \mu}(x) e_{\nu}^{\enskip a}(x){\bf
{d}}y_{\mu}{\bf {d}}x^{\nu} = \Omega_{0}^{0}(x)$.  This condition will be
refined further below.

Summarizing the proof so far, we have
$$\eqalign{
\omega_{b}^{a} &= \omega_{b}^{a}({\bf{e}}(x)) \cr
\omega_{0}^{a} &= {\bf {e}}^{a}(x)  \cr
\omega_{a}^{0} &= {\bf {f}}_{a} +  {\cal {R}}_{ab}({\bf{e}}(x))  {\bf
{e}}^{b} \cr
\omega_{0}^{0} &= 0  \cr}$$
with curvatures satisfying
$$\eqalign{
\Omega_{b}^{a} &= {\bf {C}}_{b}^{a}({\bf{e}}(x))  -  \Delta_{bc}^{da}{\bf
{f}}_{d}{\bf {e}}^{c} \cr
\Omega_{0}^{a} &= 0 \cr
\Omega_{0}^{0} &= {\bf {f}}_{a}{\bf {e}}^{a} \cr} $$
Finally we use eq.(4.1c).  Expanding the co-torsion in the $({\bf {e}}^{a},
{\bf {f}}_{a})$ basis we have
$${\bf {df}}_{a} + {\bf {d}}{\cal {R}}_{a} = \omega_{a}^{b}{\bf {f}}_{b} +
\omega_{a}^{b}{\cal {R}}_{b}+ {\scriptstyle {{1 \over 2}}}
\Omega_{abc}^{0}{\bf {e}}^{b}{\bf {e}}^{c} +
\Omega_{ac}^{0b}{\bf {f}}_{b}{\bf {e}}^{c} \eqno(5.10) \hskip .5in $$
Since ${\bf {e}}^{a}$ and ${\bf {f}}_{a}$ are independent, eq.(5.10)
separates into two
parts, namely
$$\eqalignno{
{\bf {d}}{\cal {R}}_{a} &=  \omega_{a}^{b}{\cal {R}}_{b}+ {\scriptstyle {{1
\over 2}}}\Omega_{abc}^{0}{\bf {e}}^{b}{\bf {e}}^{c} &(5.11) \hskip .5in \cr
{\bf {df}}_{a} &= \omega_{a}^{b}{\bf {f}}_{b} +  \Omega_{ac}^{0b}{\bf
{f}}_{b}{\bf {e}}^{c} &(5.12) \hskip .5in  \cr } $$
The consistency of the separation follows because all of the coefficients
of these equations
are functions of $x$ only.  This means that the Bianchi identity of
eq.(5.10) automatically implies both of the Bianchi identities following
from eqs.(5.11) and (5.12) and {\it {vice versa}}.  Not only do these
equations fix the remaining outstanding curvatures in terms of ${\bf
{e}}^{a}$ and ${\bf {f}}_{a}$, but eq.(5.12) also shows that ${\bf
{f}}_{a}$ is in involution.  Once again invoking the Frobenius theorem, we
see that the biconformal bundle is foliated by a second set of
11-dimensional sub-bundles with Weyl group fibers and base manifold spanned
by ${\bf {d}}x^{\mu}$.  Thus, there are two foliations of the biconformal
bundle by 11-dimensional sub-bundles such that the 4-dimensional base space
for each is the space of leaves of the other.  Since the Frobenius theorem
guarantees that each base space is a manifold, we may conclude that both
spaces of leaves are manifolds, hence regular.  In particular, the torsion
foliation is regular.

On the ${\bf {f}}_{a} = 0$ manifold we have the structure equations
$$\eqalignno{
{\scriptstyle { {1 \over 2} }} C_{\enskip bcd}^{a}{\bf {e}}^{c}{\bf
{e}}^{d} &= {\bf {d}}\omega_{b}^{a} - \omega_{b}^{c} \omega_{c}^{a} -
\Delta_{bd}^{ha} {\cal {R}}_{hc} {\bf {e}}^{c} {\bf {e}}^{d} &(5.13a)
\hskip .5in \cr
{\bf {de}}^{a} &= {\bf {e}}^{b}\omega_{b}^{a} &(5.13b) \hskip .5in  \cr
\omega_{a}^{0}\vert_{y=const.} &=  {\cal {R}}_{ab} {\bf {e}}^{b} &(5.13c)
\hskip .5in \cr
\Omega_{0}^{0} &= 0 = \omega_{0}^{0} &(5.13d) \hskip .5in \cr }$$
Eqs.(5.13) are equivalent to eqs.(2.1) for a Riemannian geometry.

Finally, returning to the torsion involution, at each fixed value of
$x^{\mu}$ we have the
single remaining structure equation
$${\bf {df}}_{a} = 0 \eqno(5.14) \hskip .5in  $$
with all other connection forms vanishing.  Therefore ${\bf {f}}_{a}$ spans
a flat 4-dim space.  From the local Lorentz action on the biconformal
fibers we see that the Lorentz group acts on this space, so even though we
have not introduced a notion of orthogonality on the 4-dimensional
subspaces, we can identify it with Minkowski space, up to global topology.
We will assume trivial topology for this space, although it is simple to
generalize to topologies which are quotients of Minkowski space by discrete
subgroups\footnote{$^{1}$}{{\smallrm {In these non-trivial cases the vector
addition must be made modular and the isomorphism established in the
remainder of the proof is with the corresponding non-simply connected
Minkowski space.  In all other regards the proof is the same.}}}.  In
either case, there exists such a vector space at each point of the
spacetime, so the $\omega_{0}^{0} = 0$ slice of the biconformal bundle is
the local direct product of a 4-dimensional Riemannian spacetime with the
direct sum of vector spaces isomorphic to the Weyl group and Minkowski
vector space (or a discrete quotient of Minkowski space).

To conclude the proof we establish a particular vector space isomorphism
between the Minkowski space (now assumed simply connected) spanned by ${\bf
{f}}_{a}$ and the tangent space spanned by ${\partial \over \partial
x^{a}}$.  First notice that such a mapping behaves correctly under Lorentz
transformations.  The reason is that ${\bf {d}}y_{a}$ and ${\bf {d}}x^{a}$
have opposite Lorentz transformation and dilation properties, as seen from
eqs.(3.8b,c):
$$\eqalignno{
{\hat{\omega}}_{0}^{a} &= \Lambda^{a}_{\enskip b} \> \omega_{0}^{b}\>
{\bar{\Lambda}}^{0}_{\enskip 0}  &(3.8b) \hskip .5in \cr
{\hat{\omega}}_{a}^{0} &=  \Lambda^{0}_{\enskip 0} \> \omega_{b}^{0}\>
{\bar{\Lambda}}^{b}_{\enskip a} &(3.8c) \hskip .5in \cr} $$
Therefore, since ${\bf {e}}^{a}$ spans the co-tangent space, ${\bf
{f}}_{a}$ transforms
in the same way as a tangent basis.

To arrive at the appropriate mapping for the isomorphism between the
Minkowski space spanned by ${\bf {f}}_{a}$ and the tangent space we use
$$\eqalignno{
0 = {\bf {d}}\Omega_{0}^{0} &= {\bf {d}}({\bf {f}}_{a}{\bf {e}}^{a})  \cr
&= \partial_{\alpha} f_{\mu}^{\enskip \nu}{\bf {d}}x^{\alpha}{\bf
{d}}y_{\nu}{\bf {d}}x^{\mu}   &(5.15) \hskip .5in  \cr } $$
where $f_{\mu}^{\enskip \nu} \equiv f_{a}^{\enskip \nu} e_{\mu}^{\enskip
a}$.  Since $f_{\mu}^{\enskip \nu}$ is a function of $x$ only this may be
written as ${\bf {df}}^{\nu} = 0$, and we can find a chart on which ${\bf
{f}}^{\nu} = {\bf {d}}z^{\nu}$ for some $z^{\nu}(x)$.  This puts the
symplectic dilational curvature into the Darboux form
$$\Omega_{0}^{0} = {\bf {d}}y_{\mu}{\bf {d}}z^{\mu} \eqno(5.16) \hskip .5in $$
This result is actually slightly stronger than the usual Darboux theorem,
since ${\bf {d}}z^{\mu}$ is required to span the subspace of the torsion
involution.

Of course, in transforming to $z^{\mu}$ as the spacetime coordinate we have
changed the components of the solder form, which may be written as
$${\bf {e}}^{a} = e_{\mu}^{\enskip a} {\bf {d}}x^{\mu} = e_{\mu}^{\enskip
a} {\partial x^{\mu} \over \partial z^{\nu}} {\bf {d}}z^{\nu} \equiv
\tilde{e}_{\nu}^{\enskip a}(x(z)){\bf {d}}z^{\nu}  \eqno(5.17a) \hskip .5in
$$
At the same time the components of the co-solder form are altered according to
$$\eqalignno{
{\bf {f}}_{a} &= f_{a}^{\enskip \mu}{\bf {d}}y_{\mu} \cr
&= f_{\mu}^{\enskip \nu} e_{a}^{\enskip \mu} {\bf {d}}y_{\nu} \cr
&={\partial x^{\nu} \over \partial z^{\mu}} e_{a}^{\enskip \mu} {\bf
{d}}y_{\nu} \cr
&= \tilde{e}_{a}^{\enskip \nu}(x(z)) {\bf {d}}y_{\nu} &(5.17b) \hskip .5in
\cr } $$
so that the co-solder form is determined entirely by the solder form.

With the solder and co-solder forms related as in eqs.(5.17), the entire
biconformal connection is fully determined by the 4-dimensional spacetime
solder form.  Moreover, we now have a uniquely specified mapping between
the co-basis of the biconformal space and a basis in the tangent space,
which preserves the usual dual-basis relationship between tangent and
co-tangent bases.  For if we have the dual bases
$$ e_{a}  \longleftrightarrow {\bf {e}}^{a}  \eqno(5.18) \hskip .5in $$
then we can set
$$\phi: {\bf {f}} _{a}  \longrightarrow e_{a} \eqno(5.19) \hskip .5in $$
or for the coordinates
$$\phi: {\bf {d}}y_{\mu} \longrightarrow {\partial \over \partial z^{\mu}}
\eqno(5.19)
\hskip .5in  $$
where $y_{\mu}$ and $z^{\nu}$ are conjugate with respect to the symplectic
form $\Omega_{0}^{0}$.  Clearly, the mapping of the bases establishes the
isomorphism between the vector spaces, concluding the proof.
\bigskip
\leftskip=0in
\rightskip=0in

Theorem I establishes a 1-1 relation between the class of $\omega_{0}^{0} =
0$ sub-bundles of normal biconformal spaces and the class of 4-dimensional
Riemannian
geometries, since the converse obviously holds $-$ given any Riemannian
geometry we can immediately construct the Whitney sum of its tangent and
orthonormal frame bundles, and invoke the isomorphism $\phi^{-1}$ on the
tangent space to provide the co-space basis.

The principal correspondence of Theorem I is immediately seen by the {\it
{standard form}} of the normal biconformal connection and curvatures, which
we collect here for future reference.  Renaming $z^{\mu} \longrightarrow
x^{\mu}$ we have connection forms
$$\eqalignno{
\omega_{b}^{a} &= \omega_{b}^{a}({\bf{e}}(x)) &(5.20a) \cr
\omega_{0}^{a} &= {\bf {e}}^{a}(x) &(5.20b)  \cr
\omega_{a}^{0} &= e_{a}^{\enskip \mu}(x) {\bf {d}}y_{\mu} +  {\cal
{R}}_{ab}({\bf{e}}(x))  {\bf {e}}^{b} &(5.20c) \cr
\omega_{0}^{0} &= 0 &(5.20d)  \cr}$$
which lead to the curvatures
$$\eqalignno{
\Omega_{b}^{a} &= {\bf {C}}_{b}^{a}({\bf{e}}(x))  +  \Delta_{bc}^{da}{\bf
{f}}_{d}{\bf {e}}^{c} &(5.21a) \cr
\Omega_{0}^{a} &= 0 &(5.21b) \cr
{\scriptstyle {{1 \over 2}}} \Omega_{abc}^{0} \> {\bf {e}}^{b}{\bf {e}}^{c}
&=  {\bf {d}}{\cal {R}}_{a} - \omega_{a}^{b}{\cal {R}}_{b} \equiv  {\bf {D}}{\cal {R}}_{a} &(5.21c) \cr\Omega_{ac}^{0b}\>{\bf {f}}_{b}{\bf {e}}^{c} &= {\bf
{df}}_{a} - \omega_{a}^{b}{\bf {f}}_{b} \equiv  {\bf {Df}}_{a} &(5.21d) \cr
\Omega_{0a}^{0b}\>{\bf {f}}_{b}{\bf {e}}^{a} &= \delta_{a}^{b}{\bf
{f}}_{b}{\bf {e}}^{a} &(5.21e) \cr} $$
Henceforth, we will speak of a Riemannian geometry and its corresponding
normal
biconformal space interchangeably.

Notice that the isomorphism $\phi$ developed in the proof of the Theorem
establishes for the entire class of normal biconformal spaces the
identification of the
co-space with momentum-like variables.  In the simple one-particle picture
of references [31, 33], the $y$-coordinates are directly proportional to
the components of the particle momentum.  In the general field theories
studied here, we see that ${\bf {d}}y_{\mu}$ is proportional to the
generator of infinitesimal translations.

The results of Theorem I are valid as long as we restrict the Lorentz
transformations of biconformal space to the usual $x$-dependent ones of
Riemannian geometry and the rescalings to the usual constant ones.  We can
avoid the restriction against $x$-dependent scalings as well as the
restriction to the $\omega_{0}^{0} = 0$ sub-bundle by extending the
homeomorphism in the following way.
\medskip
\leftskip=.7in
\rightskip=.25in
\parindent=-.7in
{\it {Corollary 1:}} A normal biconformal space with gauge transformations
restricted to $\Lambda^{a}_{\enskip b} = \Lambda^{a}_{\enskip b}(x)$ and
$\Lambda^{0}_{\enskip 0} = \Lambda^{0}_{\enskip 0}(x)$ is homeomorphic to
the triple Whitney sum bundle of the tangent bundle, the bundle of
orthonormal Lorentz frames and the scale bundle over a 4-dimensional Weyl
spacetime with vanishing dilational curvature.
\medskip
\leftskip=.5in
\parindent=-.5in
{\it {Proof:}} Off of the $\omega_{0}^{0} = 0$ sub-bundle of a normal
biconformal space, $\omega_{0}^{0}$ remains an exact form.  If we embed the
given Riemannian geometry as the ${\bf {W}}=0$ cross-section of a Weyl
geometry of vanishing dilation, then the 1-dimensional scaling fiber of the
Weyl geometry is isomorphic to the 1-dimensional fiber of $x$-dependent
gauges $\Lambda^{0}_{\enskip 0}(x)$.  This extends the homeomorphism to one
between the $x$-dependent gauge sector of the whole biconformal space and
the triple Whitney sum over a dilation-trivial 4-dimensional Weyl geometry
of the tangent, orthonormal frame and scale bundles.
\medskip
\leftskip=0in
\rightskip=0in
\parindent=.25in
We can go the final step and remove the restriction on the gauge
transformations by recalling that a {\it {Finsler geometry}} is a
differential geometry in which the metric and connection are allowed to
depend on not only the coordinates, but also on tangent vectors to curves.
Formulated as a fiber bundle, the allowed orthonormal frame transformations
of a Finsler space will depend on 4-velocities as well as the coordinates.
This is precisely the sort of transformation provided by a general Lorentz
transformation in biconformal space.  Making the obvious generalization to
a Weyl-Finsler geometry, we immediately see the full import of a normal
biconformal space.  Calling a Weyl-Finsler space {\it {trivial}} if it
permits a cross-section on which it reduces to a Riemannian spacetime, we
immediately have the following:
\medskip
\leftskip=.7in
\rightskip=.25in
\parindent=-.7in
{\it {Corollary 2:}} A normal biconformal space is homeomorphic to the
triple Whitney sum bundle of the tangent bundle, the bundle of orthonormal
Lorentz frames and the scale bundle over a trivial 4-dimensional
Weyl-Finsler spacetime.
\medskip
\leftskip=0in
\rightskip=0in
\parindent=.25in
In the remainder of this study, we will restrict our attention to the
content of the central Theorem, since our principal interest lies in
establishing the biconformal equivalent of standard results in general
relativity.  As we show in the next section, Theorem I allows us to write
the biconformal equivalent of the Einstein equation.  Three facts make the
biconformal Einstein equation simpler than the conformal Einstein equation
$-$ the separation of the Ricci and conformal parts of the curvature, the
reduction of the dimension of the fiber symmetry in biconformal space from
11 to 7, and the natural inner product.  The Ricci tensor becomes part of
the co-solder form with its curl contributing to the co-torsion, while only
the conformal curvature remains as the spacetime term of the curvature of
the spin connection.  The reduced fiber symmetry then insures that the
Ricci tensor is in fact biconformally tensorial and the inner product
allows complete isolation of the Ricci tensor.  After using these facts to
write the Einstein equation in biconformal space, we present a new
geometric model of electromagnetism using some of the additional degrees of
freedom present in more general biconformal spaces.
\bigskip
\noindent 6.  The gravitational and electromagnetic field equations in
normal biconformal space.
\medskip
The most immediate importance of Theorem I is in establishing that the
normal biconformal equivalent of the Einstein equation is simply the
Einstein equation itself.  There are three principal reasons that this
occurs.  First, the separation between the curvature and co-torsion
achieved in a normal biconformal space separates out the Ricci part of the
curvature tensor as part of the co-solder form.  Second, because the
8-dimensional base space removes the translations from the fiber symmetry
while leaving the Weyl vector to absorb rescalings, ${\cal {R}}_{a}$
transforms as a tensor even under rescalings.  Finally, the natural inner
product on biconformal spaces permits complete isolation of the Ricci
tensor.

Notice that arriving directly at the Einstein equation avoids the
troublesome issue mentioned in the introduction, namely, the cubic
dependence of the conformal equivalent of the vacuum Einstein equation on
the curvature [49].  While the biconformal conditions provided here are, of
course, not invariant under all conformal transformations, they are
invariant under local scalings.  We discuss these differences below.

To begin our study of field equations, we use the biconformal inner product
to separate out the Ricci part of curvature.  We immediately find the
following corollary.

\medskip
\parindent=.75in
\hang
\noindent{\it {Corollary 3:}}  The Riemannian geometry of a normal
biconformal space is
Ricci flat if and only if the trace-free basis is orthonormal.
\medskip

\parindent=.5in
\rightskip=.5in
\hang
\noindent{\it {Proof:}} The inner product in the trace-free basis is
$$\eqalignno{
\langle {\bf {e}}^{a}, {\bf {e}}^{b} \rangle  &= \langle \omega_{0}^{a},
\omega_{0}^{b} \rangle &\cr
&= 0 &(6.2a) \hskip .5in  \cr
\langle {\bf {e}}^{a}, {\bf {f}}_{b} \rangle  &= \langle \omega_{0}^{a},
\omega_{b}^{0} - {\cal {R}}_{bc}{\bf {e}}^{c} \rangle \cr
&= \langle \omega_{0}^{a}, \omega_{b}^{0} - {\cal {R}}_{bc}\omega_{0}^{c}
\rangle
\cr
&= \delta_{b}^{a} &(6.2b) \hskip .5in  \cr
\langle {\bf {f}}_{a}, {\bf {f}}_{b} \rangle  &= \langle \omega_{a}^{0} -
{\cal
{R}}_{ac}\omega_{0}^{c}, \omega_{b}^{0} - {\cal {R}}_{bd}\omega_{0}^{d}
\rangle \cr
&= -{\cal {R}}_{ab}-{\cal {R}}_{ba} &(6.2c) \hskip .5in  \cr }$$
which is orthonormal if and only if
$${\cal {R}}_{ab}=0, \eqno(6.3) \hskip .5in $$
Eq.(6.3) immediately implies Ricci-flatness.
\medskip
\rightskip=0in
\parindent=.25in
Therefore, Ricci flatness arises in biconformal space as an orthonormality
condition on the standard normal biconformal basis.  Indeed, the Ricci
tensor is a direct measure of the degree to which the co-solder form fails
to be null.

By treating the co-solder projection $B_{ab}$ as initially independent of
the curvature, we can use the tracelessness condition to impose the full
Einstein equation including phenomenological sources for the gravitational
field.  Specifically, we let $B_{ab}$ be identified with an appropriately
trace-altered form of the stress-energy tensor:
$$\eqalignno{
B_{ab} &\equiv {\cal {T}}_{ab} \equiv - {1 \over 2}(T_{ab} - {1 \over 3}
\eta_{ab} T) &(6.4a) \cr
{\cal {T}}_{a} &\equiv {\cal {T}}_{ab} {\bf {e}}^{b} &(6.4b) \cr } $$
where $T_{ab}$ is the stress-energy tensor and T is its trace.  Notice that
the (geometric) units for $T_{ab}$ are correct for this identification to
scale properly.  Then it follows immediately from the tracelessness
condition, eq.(5.6), and eqs.(6.4) that
$${\cal {R}}_{a} = {\cal {T}}_{a} \eqno(6.5) $$
Eq.(6.5) is simply the Einstein equation with the trace term partitioned
differently between the Ricci and stress-energy tensors.  It appears that
the extra four dimensions and the resulting co-torsion fields represent
various aspects of matter.  We shall see this pattern continue when we
consider a new geometric model for electromagnetism below.

Recalling the remarks concerning gauging in Sec.(3), we see that under a
local Lorentz transformation $\Lambda^{b}_{\enskip a}$ and a local scale
change $e^{\phi}$, eq.(6.3) simply changes to
$$e^{\phi} {\cal {R}}_{b}{\bar{\Lambda}}^{b}_{\enskip a} = 0 \eqno(6.6) $$
and so Corollary 3 remains valid in any Lorentz-Weyl gauge.  Similarly,
eq.(6.5) continues to hold.  Neither of these equations would be invariant
under the conformal or inhomogeneous Weyl group, since the inverse
translations would then change the co-solder form inhomogeneously.
Nonetheless, unlike the Ricci tensor of a Riemannian geometry, the presence
of the Weyl form (required to be exact but not necessarily zero in a normal
biconformal space) keeps the scale transformation of ${\cal {R}}_{a}$
tensorial.  As might be expected, the inhomogeneous transformation of the
Ricci tensor resulting from inverse translations shows up instead in the
coordinate transformations of the extra 4-dimensions.  For example, if we
define a new $y$-coordinate by setting
$$y_{\mu} = z_{\mu} + h_{\mu}(x) \eqno(6.6) $$
then eq.(5.20c) for the co-solder form becomes
$$\omega_{a}^{0} = e_{a}^{\enskip \mu}{\bf {d}}z_{\mu} +  {\cal
{R}}_{ab}({\bf{e}}(x))  {\bf {e}}^{b} + e_{a}^{\enskip \mu}{\partial
h_{\mu} \over \partial x^{\nu} } e_{b}^{\enskip \nu}{\bf {e}}^{b} $$
so ${\cal {R}}_{a}$ picks up an additive contribution.  Of course,
eqs.(6.3) and (6.5) are covariant with respect to changes of the
$x$-coordinate of the form $\bar{x}(x)$.

This situation contrasts strongly with standard conformal gaugings.  For a
4-dimensional spacetime to be conformally Ricci-flat requires the existence
of a scalar field $\phi$ with gradient $\phi_{\mu} \equiv \phi_{,\mu}$
satisfying
$${\bf {D}}{\cal {R}}_{a} - \phi_{b}{\bf {C}}_{a}^{b} =  0 \eqno(6.7) $$
By finding conditions on the Weyl curvature ${\bf {C}}_{a}^{b}$ for the
existence of a tensor field $D_{c}^{ade}$ such that $M_{c}^{b} \equiv
D_{c}^{ade} C_{ade}^{b}$ is nondegenerate, it is possible to solve this
equation for $\phi_{b}$.  The conditions on ${\bf {C}}_{a}^{b}$, which turn
out to be cubic [49], then give a test for the conformal Ricci-flatness.

The standard picture is improved somewhat if the usual connection on a
Riemannian spacetime is extended to a conformal connection.  This separates
out various conformally covariant combinations of the curvatures, notably
the Thompson-Szekeres-Yang term [49, 51-53], ${\bf {D}}{\cal {R}}_{a}$.
But the net result is still either eq.(6.7), or the reduction of the
conformal connection to a Weyl connection.  Of these two possibilities,
only the reduction to a Weyl connection allows a fairly simple
representation of the scale-invariant Einstein equation.  But in that case,
four of the conformal degrees of freedom are lost in the reduction to a
Weyl connection.

Moving to a biconformal gauging retains the maximal number of conformal
degrees of freedom while gaining the simplifying advantage of a Weyl
connection. Because normal biconformal spaces are essentially Weyl
geometries, the Weyl vector takes care of the allowed scale transformations
without requiring the solution of a more complicated set of equations.  The
elimination the inverse translations by using an 8-dimensional base
manifold has removed those transformations which interfere with the
invariance of the Ricci tensor, while retaining the increased number of
transformations as coordinate instead of gauge degrees of freedom.  As we
have shown, it then remains possible to recover the usual Riemannian
geometries of general relativity and the Einstein equation in a
biconformally invariant way.

Next we extend slightly beyond the constraints of normal biconformal spaces
to look at a new geometric model for electromagnetism.  Returning to the
orthonormality condition of the corollary, it is evident that the
$(\omega_{0}^{a}, \omega^{0}_{a})$ basis becomes identical to the $({\bf
{e}}^{a}, {\bf {f}}_{a}) $ basis when ${\cal {R}}_{ab} = 0$.  This raises
the question of what class of ${\bf {e}}$-co-frames is orthogonal.  The
answer follows immediately from eqs.(6.2) with ${\cal {R}}_{ab}$ replaced
with a general matrix $B_{ab}$.  Clearly, changing the antisymmetric part
of $B_{ab}$ has no effect on orthonormality.  In order to allow $B_{ab}$ to
have a nonvanishing antisymmetric part, we can replace the normality
conditions of our central theorem by slightly weakening the tracelessness
condition, since it is the tracelessness condition that forces $B_{ab} =
{\cal {R}}_{ab} = B_{ba}$.  Instead of
$$\Omega_{bac}^{a} = 0$$
we now require only the vanishing of the symmetric part,
$$\Omega_{bac}^{a} + \Omega_{cab}^{a} = 0 \eqno(6.8) $$
This has the desired effect.  Writing
$$\omega_{a}^{0} = {\bf {f}}_{a} + (S_{ab} + F_{ab}){\bf {e}}^{b} \eqno(6.9) $$
where $S_{ab} = S_{ba}$ and $ F_{ab} = -F_{ba}$, and imposing eq.(6.8) we
easily
find that $F_{ab}$ is undetermined while $S_{ab} = {\cal {R}}_{ab}$.

We must now check the effect of this change on the other structure
equations.  The
remaining equations are
$$\eqalignno{
{\bf {de}}^{a} &=  {\bf {e}}^{b} \omega_{b}^{a}  &(6.10a) \cr
{\bf {d}}\omega_{a}^{0} &= \omega_{a}^{b}\omega_{b}^{0}+ \Omega_{a}^{0}
&(6.10b)  \cr
{\bf {d}}\omega_{0}^{0} &= 0 = {\bf {e}}^{a} \omega_{a}^{0}+ \Omega_{0}^{0} &
(6.10c) \cr}$$
Eq.(6.10a) has no effect on $F_{ab}$, while the exterior derivative of
eq.(6.10c) still implies ${\bf {e}}^{a} \Omega_{a}^{0} = 0$ and thus
$\Omega_{a}^{0bc} = 0$.  This, in turn, allows us to write $f_{a}^{\enskip
\mu} = e_{a}^{\enskip \mu}$ as before.  Now when we combine eqs.(6.9) and
(6.10c) and impose ${\bf {d}} \Omega_{0}^{0} = 0$ we find
$${\bf {dF}} = -{\bf {d}}\Omega_{0}^{0} - {\bf {d}}({\bf {e}}^{a} {\bf
{f}}_{a}) =
0  \eqno(6.11) $$
and thence
$$\eqalignno{
F_{[\alpha \beta, \mu]} &= 0 &(6.12a) \cr
F_{\alpha \beta}^{\enskip \enskip, \mu} &= 0 &(6.12b) \cr } $$
After seeing from eq.(6.12b) that $F_{ab}$ is independent of $y_{\mu}$,
eq.(6.12a) shows that it arises from a potential.  Therefore, ${\bf {F}}$
is a closed 2-form on spacetime.

Finally, we consider the effect of eq.(6.10b).  Again substituting eq.(6.9)
and using the
independence of $F_{ab}$ from $y_{\mu}$ to separate off the ${\bf {f}}_{a}$
terms, we
are led to
$${\bf {DF}}_{a} + {\bf {D}}{\cal {R}}_{a} = {\scriptstyle {{1 \over 2}}}
\Omega^{0}_{abc}{\bf {e}}^{b}{\bf {e}}^{c} \eqno(6.13) $$
so we find that the source for ${\bf {F}}$, like that for the gravitational
field, resides in the
spacetime co-torsion.

We can carry this result further if we use the usual spacetime metric
constructed from the
solder form,
$$g_{\alpha \beta} \equiv \eta_{ab} e_{\alpha}^{\enskip a}
e_{\beta}^{\enskip b}
\eqno(6.14) $$
and its inverse $g^{\alpha \beta}$ to contract the spacetime co-torsion.
In components with the usual 4-dimensional notation we start with
$$ \Omega^{0}_{\alpha \beta \mu} =  F_{\alpha \beta; \mu} - F_{\alpha \mu;
\beta}+ {\cal
{R}}_{\alpha \beta; \mu} -  {\cal {R}}_{\alpha \mu; \beta}  \eqno(6.15) $$
Contracting on $(\alpha, \beta)$ we find
$$g^{\alpha \beta} \Omega^{0}_{\alpha \beta \mu} = - F^{\beta}_{\enskip
\mu; \beta} +
{\cal {R}}_{; \mu} - {\cal {R}}^{\beta}_{\enskip \mu; \beta} \eqno(6.16) $$
Then using the contracted Bianchi identity for the Einstein tensor we
easily find ${\cal
{R}}_{; \mu} - {\cal {R}}^{\beta}_{\enskip \mu; \beta} = 0$ so the
electromagnetic source neatly separates from the gravitational source as
$$J_{\mu} \equiv - g^{\alpha \beta} \Omega^{0}_{\alpha \beta \mu} =
F^{\beta}_{\enskip \mu; \beta} \eqno(6.17) $$

This geometric theory of electromagnetism is quite distinct from previous
unifications with gravity, including those based on scale-invariant
theories.  In Weyl's original scale-invariant theory of electromagnetism
[1], the Weyl vector is identified with the electromagnetic potential.
Though Weyl's theory fails to agree with experiment, its biconformal
version [33] provides a satisfactory unification.  But the theory presented
here makes an entirely different postulate from either [1] or [33] for the
identification of the electromagnetic field among the geometric variables.
In the model of this section, the electromagnetic field strength is
identified with part of the co-solder form, and its source is identified
with a well-defined trace of the co-torsion.  The fact that the field
arises from a potential is then a consequence of the biconformal structure
equations.  In this new model, and in direct contrast to previous
scale-invariant EM-gravity unifications, the Weyl vector vanishes.

It is interesting to speculate that if we let the Weyl vector be non-zero
in a biconformal space there would actually be two independent EM-like
fields.  Perhaps some combination of these can be successfully identified
with the photon and $Z^{0}$ pair of electroweak fields.  There is clearly
enough internal symmetry in biconformal spaces for such a model.  Not only
does the extra co-space allow a set of internal transformations, but there
is also a 4-parameter class of invariant tensors.  In addition, general
biconformal spaces have far more fields than the normal biconformal spaces
studied here.  However, a great deal more research is required to determine
how to identify the fields correctly and to check the consistency of the
resulting models.
\bigskip
\noindent 7.  Summary
\medskip
We have found the necessary and sufficient conditions on an 8-dimensional
gauging of the 15-dimensional conformal group for the $\omega_{0}^{0}=0$
slice of the resulting normal biconformal space to be homeomorphic to the
Whitney sum bundle of the tangent and orthonormal frame bundles of a
4-dimensional Riemannian geometry.  Our central theorem provides for the
immediate unique extension of a general Riemannian spacetime to a normal
biconformal space.

In addition, we have written the Einstein equation for general relativity
in a scale-invariant form using the Weyl structure of the biconformal
bundle.  Also, by slightly weakening one of the normality conditions we
provide a new geometric candidate for the electromagnetic field and show
how to specify the biconformal fields to enforce the Maxwell equations.
The model therefore provides a scale-invariant geometric unification of
general relativity and Maxwell electromagnetism.

\vfil

The author wishes to thank Y. H. Clifton for many helpful discussions and
C. G. Torre and A. Wehner for their careful readings of the manuscript.

\vfil
\break
\bigskip
\bigskip
\hfil References
\medskip
\item{[1]} Weyl, H., Sitzung. d. Preuss. Akad. d. Wissensch. (1918) 465.
Reprinted in:
The Principle of Relativity, Chapter XI, (Dover, 1923) 199 -216.
\item{[2]} Weyl, H., {\it {Space-Time-Matter}}, Dover Publications, New
York (1952).
Originally:  H. Weyl,  Raum-Zeit-Materie, (3rd Ed.,1920), Chapts II \& IV,
§§34+35,
p.242,et seq.
\item{[3]} Weyl, H., Math. Zeitschr.  2 (1918) 384; Ann. d. Physik,  54
pp117 (1918); Ann. d. Physik,  59 (1919) 101.
\item{[4]} Weyl, H., Phys. Zeitschr. 21 (1920) 649; Ann. d. Phys. 65
541(1921); Phys. Zeitschr. 22 (1921) 473.
\item{[5]} Weyl, H., Relativity Symposium, Nature 106 781(1921); Zeit. f.
Physik, 56 (1929) 330.
\item{[6]} Pauli, W., {\it {Theory of Relativity}} Translated by G. Field,
Dover Press,
New York (1958) 192
\item{[7]} A. Einstein, S.B. preuss. Akad. Wiss. 478 (1918), including
Weyl's reply.
\item{[8]} F. London,  Zeitschr. f. Physik 42 375 (1927).
\item{[9]} P.A.M.Dirac, Proceedings of the Royal Society, A209, 291 (1951).
\item{[10]} P.A.M.Dirac, Proceedings of the Royal Society, A212, 330 (1952).
\item{[11]} Pauli, W., {\it {Theory of Relativity}} Translated by G. Field,
Dover Press,
New York (1958) 192
\item{[12]} Adler, R., M. Bazin, M. Schiffer, {\it {Introduction to General
Relativity}},
McGraw Hill, New York (1965) 401.
\item{[13]} J. Ehlers, A.E.Pirani, and A.Schild, in General Relativity,
edited by L.
O'Raifeartaigh (Oxford University, Oxford, 1972).
\item{[15]} Utiyama, R., Prog. Theor. Phys. {\bf {50}} (1973) 2080.
\item{[16]} P.A.M. Dirac, Proc. Roy. Soc. London  A333 (1973) 403.
\item{[17]} Freund, P. G. O., Ann. Phys. {\bf {84}} (1974) 440.
\item{[18]} Utiyama, R., Prog. Theor. Phys. {\bf {53}} (1975) 565.
\item{[19]} P.G. Bergmann, {\it {Introduction to the Theory of
Relativity}}, Chap XVI
(Dover, 1976).
\item{[20]} Hayashi, K., M. Kasuya and T. Shirafuji, Prog. Theor. Phys.
{\bf {57}}
(1977) 431.
\item{[21]} Hayashi, K. and T. Kugo, Prog. Theor. Phys. {\bf {61}} (1979) 339.
\item{[22]} J. Audretsch,  Phys. Rev. D27 2872 (1983).
\item{[23]} J. Audretsch, F. Gähler and N. Straumann,  Commun. Math. Phys.
95, 41
(1984).
\item{[24]} Ranganathan, D. , J. Math Phys. {\bf {28}} (1986) 2437.
\item{[25]} Cheng, H. Phys. Rev. Lett. {\bf {61}} (1988) 2182.
\item{[26]} Caianiello, E. R., M. Gasperini, E. Predazzi and G. Scarpetta,
Phys. Lett.
{\bf {132A}}, 2, (1988) 82
\item{[27]}  Caianiello, E. R., A. Feoli, M. Gasperini and G. Scarpetta,
Intl. J.Theor.
Phys., {\bf {29}}, 2, (1990), 131 and references therein.
\item{[28]} Castro, C., Univ. of Texas and Austin, 	Center for Particle
Theory preprint, Nov. (1987).
\item{[29]} Wood, W. R. and G. Papini, Phys. Rev. D, {\bf {45}}, 10 (1991)
3617.
\item{[30]} Wheeler, J. T., Phys Rev D{\bf {41}} (1990) 431.
\item{[31]}Wheeler, J. T., to be published in  the proceedings of the
Seventh Marcel
Grossman conference, gr-qc 9411030.
\item{[32]} Wheeler, J. T., Phys Rev D{\bf {44}} (1991) 1769.
\item{[33]} Wheeler, J. T., submitted for publication.
\item{[34]} Cartan, \'{E}., La th\'{e}orie des groupes finis et continus et la
g\'{e}om\'{e}trie diff\'{e}rentielle, Paris, Gauthier-Villars (1937); see
also Misner, C. W., K. S. Thorne and J. A. Wheeler,  $Gravitation$, W. H.
Freeman and Co., San Francisco
(1970) and references therein.
\item{[35]} Klein, F., ``Erlangerprogram:  Vergleichende Betrachtungen
\"{u}ber neuere
geometrischen Forschungen" (1872), translated in Bull. Amer. Math. Soc. 2
(1893) 215-
249.
\item{[36]} Utiyama, R., Phys. Rev. {\bf {101}}, (1956) 1597.
\item{[37]} Kibble, T.W.B., J. Math. Phys., {\bf {2}}, (1961) 212.
\item{[38]} S.W. MacDowell and F. Mansouri, Phys. Rev. Lett. 38 (1977) 739.
\item{[39]} Freund, P.G.O., {\it {Introduction to Supersymmetry}}, Cambridge
University Press, Cambridge.  See Chapter 21, pp 99-105.
\item{[40]} Ferber, A. and P.G.O. Freund, Nucl. Phys. {\bf {B122}} (1977) 170.
\item{[41]} Crispim-Romao, J., A. Ferber and P.G.O. Freund, Nucl. Phys. {\bf
{B126}} (1977) 429.
\item{[42]} Kaku, M., P.K. Townsend and P. Van Nieuwenhuizen, Phys. Lett. {\bf
{69B}} (1977) 304.
\item{[43]} F. Mansouri, Phys. Rev. Lett. 42 (1979) 1021.
\item{[44]} F. Mansouri and C. Schaer, Phys. Lett. 101B (1981) 51.
\item{[45]}  Clifton, Y. H., private communication.
\item{[46]} Eguchi, T., P. B. Gilkey \& A. J. Hanson, Phys. Rep. 66, No. 6
(1980) 213.
\item{[47]} Penrose, R. and W. Rindler, {\it {Spinors and space-time}},
Vol. 2, Ch. 9, Cambridge University Press, (1986).
\item{[48]} Hughston, L.P. and T.R.Hurd Phys. Rep {\bf {100}} (1983) 273.
\item {[49]} Szekeres, P., Proc. Roy. Soc., Londoon {\bf {A274}} (1963) 206.
\item {[50]} Abrahams, R. and J. E. Marsden, {\it {Foundations of
Mechanics}}, Addison-Wesley (1978) 167.
\item {[51]} Kundt, W. and A. Thompson, Comptes Rendus Acad. Sci., Paris,
{\bf {254}}, (1962) 4257.
\item {[52]} Thompson, A., {\it {Thesis}}, London University (1962).
\item {[53]} Yang, C.N., Phys. Rev. Lett. 33, no. 7, (1974) 445.

\bye